\documentclass[structabstract]{aa}  
\usepackage{graphicx}
\usepackage{txfonts}
\usepackage{natbib}
\usepackage{longtable}
\begin{document}

\title{The remarkable solar twin HIP 56948: a prime target in the quest for other Earths
\thanks{Based on observations obtained at the
W.M. Keck Observatory, which is operated jointly by the California
Institute of Technology, the University of California and the NASA.
This paper also includes data taken 
at the McDonald Observatory of the University of Texas at Austin and
with the ESO Very Large Telescope at Paranal Observatory, Chile (observing program 083.D-0871).}
}
\titlerunning{The best solar twin}

\newcommand{\teff}{$T_{\rm eff}$ }
\newcommand{\tsin}{$T_{\rm eff}$}
\newcommand{\tef}{T_\mathrm{eff}}
\newcommand{\logg}{\log g}
\newcommand{\feh}{\mathrm{[Fe/H]}}

\author{
Jorge Mel\'endez\inst{1} \and
Maria Bergemann\inst{2} \and
Judith G. Cohen\inst{3} \and
Michael Endl\inst{4} \and
Amanda I. Karakas\inst{5} \and
Iv\'an Ram{\'{\i}}rez\inst{4,6} \and
William D. Cochran\inst{4} \and
David Yong\inst{5} \and
Phillip J. MacQueen\inst{4} \and
Chiaki Kobayashi\inst{5}\thanks{now at the Centre for Astrophysics Research, 
University of Hertfordshire, Hatfield, AL10 9AB, UK.} 
\and
Martin Asplund\inst{5}
}


\institute{
Departamento de Astronomia do IAG/USP, Universidade de S\~ao Paulo, Rua do Mat\~ao 1226, Cidade Universit\'aria, 
05508-900 S\~ao Paulo, SP, Brazil. e-mail:  jorge@astro.iag.usp.br     
\and
Max Planck Institute for Astrophysics, Postfach 1317, 85741 Garching, Germany 
\and
Palomar Observatory, Mail Stop 105-24,
California Institute of Technology, Pasadena, California 91125, USA
\and
McDonald Observatory, The University of Texas at Austin, Austin, TX 78712, USA
\and
Research School of Astronomy and Astrophysics,
The Australian National University, Cotter Road, Weston, ACT 2611, Australia
\and
The Observatories of the Carnegie Institution for Science, 813 Santa Barbara Street, Pasadena, CA 91101, USA
}

\date{Received ...; accepted ...}

 
  \abstract
   {The~Sun shows abundance anomalies relative to most solar twins. 
    If the abundance peculiarities are due to the formation
    of inner rocky planets, that would mean that only a small fraction 
    of solar type stars may host terrestrial planets.
   }
   {In this work we study HIP~56948, the best solar~twin known to date, 
to determine with an unparalleled precision how similar is to the Sun in its physical properties, 
chemical composition and planet architecture.
We explore whether the abundances anomalies 
may be due to pollution from stellar ejecta or to terrestrial planet formation.
}
   {We perform a differential abundance analysis (both in LTE and NLTE)
   using high resolution (R~$\sim$~100,000) high S/N (600-650) Keck HIRES spectra 
   of the Sun (as reflected from the asteroid Ceres) and HIP~56948.
We use precise radial velocity data from the McDonald and Keck observatories to search for planets around this star.
   }
   {We achieve a precision of $\sigma \lesssim$ 0.003~dex for
several elements. Including errors in stellar parameters the total
uncertainty is as low as $\sigma \simeq$~0.005~dex ($1\%$), which is unprecedented 
in elemental abundance studies.
The similarities between HIP~56948 and the Sun are
astonishing. HIP~56948 is only 17$\pm$7~K hotter than the Sun,  
and log~$g$, [Fe/H] and microturbulence velocity are only
+0.02$\pm$0.02~dex, +0.02$\pm$0.01~dex and +0.01$\pm$0.01 km~s$^{-1}$ higher than solar,
respectively. 
Our precise stellar parameters and a differential
isochrone analysis shows that HIP~56948 has a mass 
of 1.02$\pm$0.02M$_\odot$ and that it is $\sim$1~Gyr younger than the Sun,
as constrained by isochrones, chromospheric activity, Li and rotation.
Both stars show a chemical abundance pattern that
differs from most solar twins, but the refractory elements 
(those with condensation temperature T$_{\rm cond}$ $\gtrsim$ 1000K)
are slightly ($\sim$0.01~dex) more depleted in the Sun than in HIP~56948. 
The trend with T$_{\rm cond}$ in differential abundances (twins $-$ HIP56948)
can be reproduced very well by adding $\sim3$ M$_{\oplus}$
of a mix of Earth and meteoritic material, to the convection zone of HIP~56948.
The element-to-element scatter of the Earth/meteoritic mix
for the case of hypothetical rocky planets around HIP~56948 is only 0.0047~dex.
From our radial velocity monitoring we find 
no indications of giant planets interior to or within the habitable zone of HIP~56948. 
}
   {We conclude that HIP~56948 is an excellent candidate to host 
a planetary system like our own, including the possible presence
of inner terrestrial planets. Its striking similarity to the Sun
and its mature age makes HIP~56948 a prime target in the quest for 
other Earths and SETI endeavors.}


\maketitle

%

\section{Introduction}

In recent years there has been an important number of studies
related to solar twins, stars which are spectroscopically
almost identical to the Sun. 
The reader is refereed to  \cite{cay96} for
a review of the early history regarding the search for solar twins,
that culminated with the identification of 18 Sco as the
closest ever solar twin \citep{por97,sou04}. More recently, new solar
twins have been identified \citep{mel06,mel07,tak07,pas08,tak09,mel09,ram09}
and HIP 56948 has demoted 18 Sco as the star that most 
closely resembles the Sun \citep{mel07,tak09}.

Solar twins are useful to calibrate the zero-point of the
temperature \citep{cas10,mel10b,ram12} and metallicity 
\citep{mel10b,cas11} scales, to better characterize
the interiors of stars like the Sun \citep{baz11},
and to identify the transport mechanisms that cause Li
depletion in the Sun \citep{don09,mel10a,bau10,den10,castro11,li12}. But
most importantly, solar twins are the perfect targets
to look for small chemical abundance anomalies that may 
have been unnoticed in previous works \citep{gus_sun08,gus10}.

The first spectroscopic study designed to exploit the advantages of 
differential abundance analysis between solar twins and 
the Sun, showed that our Sun has a peculiar chemical abundance pattern,
suggested to arise from accretion of material depleted in refractory elements 
due to the formation of terrestrial planets \citep{mel09}. 
Further studies have confirmed, with different degrees 
of accuracy, that the Sun indeed has an anomalous
surface composition \citep{ram09,ram10,gon10,
gh10}\footnote{\cite{gh10}
have contested the planet signature scenario, but further
scrutiny of their work by \cite{ram10} demonstrated that in fact
the results of Gonz\'alez Hern\'andez et al. are fully consistent with
the works of \cite{mel09} and \cite{ram09}.}.
We show in the appendix A, that the reality of these abundance anomalies is well established.
Besides the important implications for planet formation
\citep{cha10} and to explain abundance anomalies in Jupiter \citep{nor09}, 
the solar abundance peculiarities may be
relevant for modeling early stellar evolution \citep{bar10}
and for solving the solar modelling crisis when 
using low solar abundances \citep{nor09,guz10}.

Even before the deficiency of refractory elements in the solar
convection zone was discovered, \cite{cas07} investigated the effect 
of accretion of metal-poor material onto the Sun as a way to 
help solve
the solar modelling problem when a low oxygen abundance \citep{all01,asp04,ma08} is adopted, 
and they found that indeed accretion
provides some improvement, but the problem is not fully solved. 
\cite{guz10} demonstrated an improvement in the comparison between stellar models and helioseismic data 
when accretion of low-Z material is taken into account (their Fig. 9), 
albeit a full resolution of the discrepancy is not found.
More detailed modelling has been recently presented by
\cite{ser11}, who use up-to-date nuclear cross-sections and 
include accretion of metal-poor and metal-rich material,
considering a range of accreted mass and different timings for accretion.
They conclude that there is somewhat better agreement with 
helioseismology for differentiated accretion, but not complete 
agreement. Overall, models with 
metal-poor accretion improve the agreement with the helium abundance
inferred from helioseismology, while metal-rich accretion
improves both the depth of the convection zone and the sound speed profile,
with exception of a model with late accretion of 0.015 M$_\odot$ of metal-poor
material, which improves the agreement with the sound speed profile.

A detailed test of the terrestrial planet formation
hypothesis was performed by \cite{cha10}, who used the
composition of about two dozen chemical elements in 
the Earth and CM chondrites (representative of the asteroid belt).
Interestingly, \cite{cha10} showed that Earth material
alone can not fully explain the peculiar solar pattern,
but that a mix of Earth and meteoritic material gives
an excellent fit for more than 20 chemical elements.
Thus, the peculiarities in the Sun could be a signature
of both the formation of terrestrial planets and of the
asteroid belt.

An interesting alternative interpretation of the solar abundance 
anomalies, from an analysis of the solar twin M67-1164
\citep{one11}, is that the chemical peculiarities may reflect 
that the Sun was born in a massive open cluster like M67. 
Using various arguments, \cite{ada10} concludes that
the birth environment of the solar system might be a moderately large
cluster with 10$^3$-10$^4$ members.
Nevertheless, only one solar twin in M67 has been analyzed to date for high precision
chemical abundances \citep{one11}, so, more observations are urgently needed 
to verify if indeed all solar twins in M67 have the same solar abundance pattern.
Also, notice that based on a dynamical study, \cite{pic12} have shown that the 
Sun could not have been born in M67.

\cite{gus10} warned about potential systematic effects in chemical abundances
due to different lines of sights when 
the Sun is compared to the solar twins. \cite{kis11} have recently studied the line-of-sight effect using high 
resolution observations at the solar equator and at latitude 45$^\circ$. Seven key chemical elements in a 
broad range of condensation temperature were analyzed by \cite{kis11}, who show that there is no difference 
in the abundances obtained at different latitudes for both volatile (to within 0.005 dex) and refractory (to within 0.002 dex)
elements. Thus, it is very unlikely that the abundance anomalies 
seen in the Sun \citep{mel09,ram09} can be attributed to line-of-sight inclination effects.

In appendix B, we show that the Sun's chemical peculiarities 
also do not arise due to the particular reflection properties of the asteroids
employed in the analyses, as expected given that the relative reflectance of asteroids show 
mostly smooth changes over hundreds of \AA\ \citep[e.g.][]{xu95,bin96,bus02,laz04,dem09}, 
about 3 orders of magnitude wider than the narrow stellar lines
used in abundance analyses.

Only some 15\% of solar type stars appear chemically similar to the Sun \citep{mel09,ram09}
and therefore we regard the Sun to have an anomalous chemical abundance when compared to other solar twins.
Assuming that the solar abundance anomalies are due to terrestrial planet formation, 
then perhaps only these 15\% of solar type stars that are chemically similar to the 
Sun may host rocky planets. 
This is a lower limit to the amount of rocky planets 
formed around other Suns, as part of those planets may fall
into their host stars, altering thus the original abundance signature.
The Kepler mission \citep[e.g.][]{bor10} will give us the first estimate of the 
frequency of Earth-sized planets in the habitable zones of solar type dwarfs.

It is important now to find, through a detailed chemical abundance analysis,
stars which are chemically identical to the Sun and which may therefore potentially host other Earths.
HIP 56948\footnote{a.k.a. {\em Intipa Awachan},http://tierneylab.blogs.nytimes.com/?s=intipa} 
is the perfect candidate for identifying subtle chemical anomalies, 
as this star has been found to be the most similar to the Sun in
stellar parameters \citep{mel07,tak09}.
The first high precision ($\sigma \sim$0.03 dex) detailed 
abundance analysis of this star showed that HIP 56948 may be one of
the rare stars with a solar abundance pattern \citep{ram09}.
In the present work, we perform a much more refined study 
($\sigma \sim 0.005$ dex) of HIP 56948, to assess its similarity
to the Sun and to which extent it may host a planetary system like ours.

\section{Observations}

\subsection{Keck HIRES spectra for high precision abundance analysis}

HIP 56948 and the Sun (reflected light from the Ceres asteroid)
were observed with HIRES \citep{vog94} at the Keck I telescope
on May 19, 2009. Exactly the same setup was used for both HIP 56948 and Ceres, 
and the asteroid was observed immediately after HIP 56948. 
A total exposure time of 800 s was used for both HIP 56948 and the Sun,
consisting of multiple observations co-added, in order to avoid non-linearity.
We stress here that for highly accurate work an asteroid should be used
instead of the daytime skylight, as there are important variations 
in the skylight spectrum with respect to the solar spectrum \citep{gray00}.
Furthermore, asteroids are essentially point sources for typical observing conditions (seeing $>$ 0.5 arcsec), 
thus the observation and data reduction for both stars and the asteroid are performed in the same way.

A resolving power of $R \approx 10^5$ was achieved using the E4
0.4"-wide slit, accepting some light loss (seeing was $\sim$ 0.7 arcsec)
in order to achieve the necessary spectral resolution. 
For HIP 56948 the signal-to-noise level measured in continuum regions is 
about 600 per pixel at 6000 \AA, while 
it is somewhat better (S/N = 650) for Ceres, for which the predicted
magnitude at the time of the observation was V = 8.30\footnote{http://ssd.jpl.nasa.gov/horizons.cgi}, 
i.e., somewhat brighter than HIP 56948 \citep[V = 8.671$\pm$0.004,][]{ols93,ram12}.

\begin{figure}
\resizebox{\hsize}{!}{\includegraphics{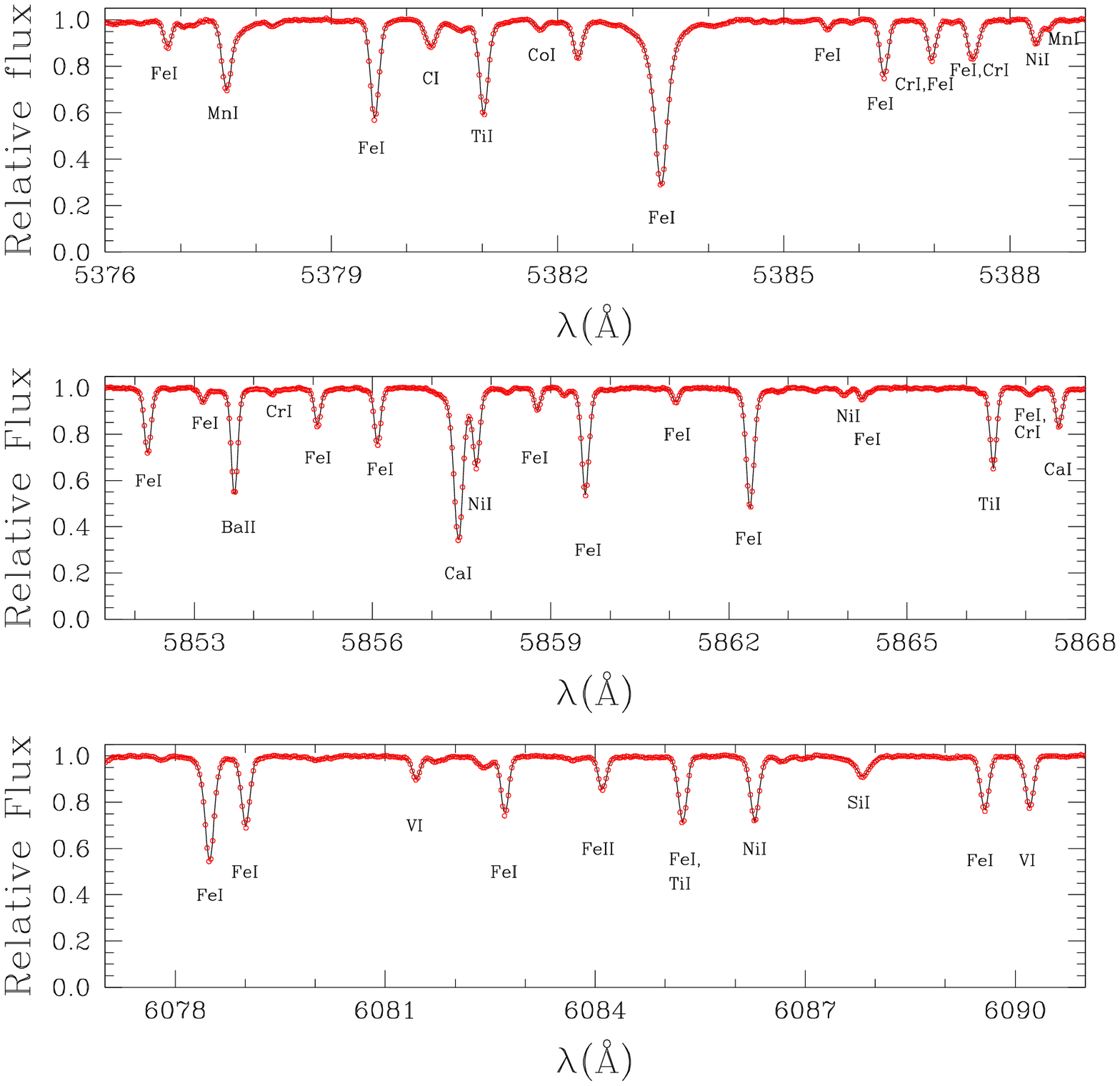}}
\caption{
Comparison of HIP 56948 (red circles) and the Sun (solid line)
in different spectral regions.
The quality of the Keck/HIRES spectra is very high for 
both HIP 56948 (S/N $\sim$ 600) and the Sun (S/N $\sim$ 650).
It is hard to distinguish any difference between both stars.
}
\label{spectra}
\end{figure}

The spectral orders were extracted  using MAKEE\footnote{MAKEE was developed 
by T. A. Barlow specifically for reduction of Keck HIRES data.
It is freely available at
http://www2.keck.hawaii.edu/realpublic/inst/hires/data\_reduction.html}.
Further data reductions (Doppler correction, continuum normalization, and 
combining spectra) were performed with IRAF.
A comparison of the reduced spectra of HIP 56948 and the Sun is shown
in Fig.~\ref{spectra}. As can be seen, the spectra are nearly indistinguishable.

\begin{figure}
\resizebox{\hsize}{!}{\includegraphics{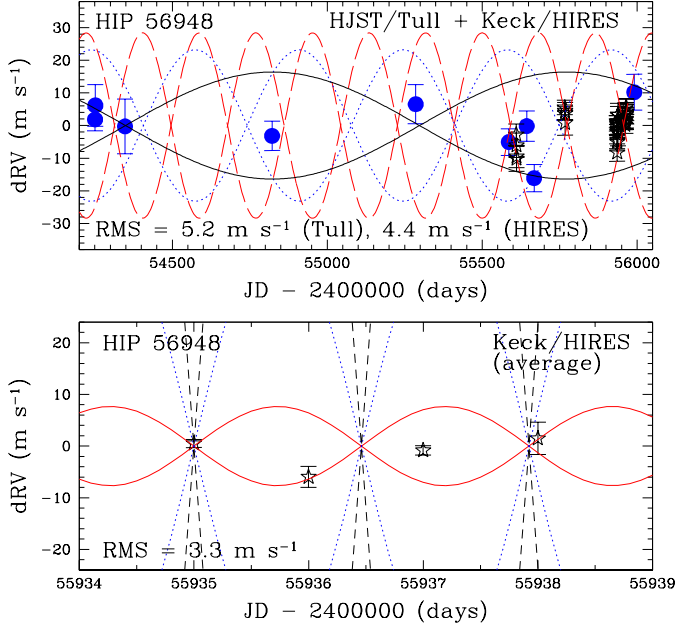}}
\caption{
{\it Upper panel}: precise radial velocities obtained with the Tull Coude Spectrograph at 
the 2.7m Harlan J. Smith Telescope (HJST) of the McDonald Observatory
(filled circles) and 
with the HIRES spectrograph at the 10m Keck telescope (stars). 
We show some circular orbits for a Jupiter-mass planet at 1 AU (long dashed line),
1.5 AU (dotted line) and 3 AU (solid line).
{\it Lower panel}: average of three short (3-min) consecutive observations taken with 
HIRES/Keck during four nights in January 2012.
Some circular orbits due to hypothetical Neptune-mass (solid line),
Saturn-mass (dotted line) and Jupiter-mass (dashed line) planets
at 0.04 AU, are shown for comparison.
No giant planets have been detected so far
in the inner regions ($<$3 AU) around HIP 56948.
}
\label{radvelmcd}
\end{figure}

\begin{figure}
\resizebox{\hsize}{!}{\includegraphics{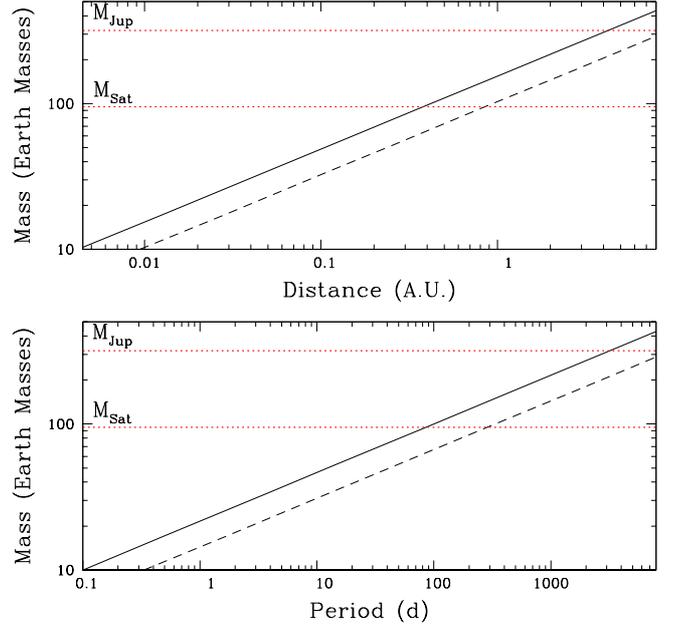}}
\caption{
Estimated sensitivity of our observations as a function of distance (upper panel)
and orbital period (lower panel). The 2-$\sigma$ and 3-$\sigma$ sensitivities are shown by dashed
and solid lines, respectively. The red dotted lines represent planets with 1 M$_{\rm Jup}$ and 1 M$_{\rm Sat}$.
Our radial velocity data discards planets as massive as Jupiter in the terrestrial planet region.
}
\label{planetsmcd}
\end{figure}

\subsection{McDonald and Keck radial velocities for planet search}

Soon after HIP 56948 was identified as the best solar twin
\citep{mel07}, the McDonald Observatory planet search program \citep[e.g.,][]{end06,rob12}
began to monitor this star.
The observations have been carried out with the
Tull Coude Spectrograph \citep{tul95} at the 2.7m Harlan J. Smith Telescope.
HIP 56948 has so far been observed nine times (from May 2007 to March 2012) and the 
scatter of the observations is 5.2 m s$^{-1}$ (discarding one outlier), 
which is consistent with the 5.1 m s$^{-1}$ median error bar.
The radial velocity data is presented in Table~\ref{tabradvel} and Fig.~\ref{radvelmcd}.

In February 2011 we started observing HIP 56948 for planets using HIRES at
Keck during time allocated to a NASA key science program to support the CoRoT mission.
\footnote{The HIRES data for HIP~56948 were obtained at times when the CoRoT field was unobservable.}
We have acquired 30 datapoints with HIRES up to February 2012.
The radial velocity data is presented in Table~\ref{tabradvelkeck} and Fig.~\ref{radvelmcd}.

For both instruments we use a temperature-controlled iodine
cell for wavelength calibration. We use the {\it Austral} code 
\citep{end00} for the computation of precise differential radial velocities.

The scatter of the HIRES observations is 4.4 m s$^{-1}$, 
which is higher than our 2.8 m s$^{-1}$ median error bar.
The difference amounts to 3.4 m s$^{-1}$ and can be explained by
typical jitter values as measured in other stars. 
For example, \cite{wri05} finds a median stellar jitter of $\sigma_{RV}' =$ 4.4 m s$^{-1}$ 
for inactive stars with (B-V) $>$ 0.60.
Interestingly, the McDonald observations seem to be less affected by
stellar jitter, due to the much longer exposure times required when
compared to the Keck observations ($\sim$3 min), 
damping the radial velocity variations due to stellar jitter.
Indeed, one of the observational strategies to reduce 
stellar noise in short time scales (mainly stellar oscillations) is 
to take relatively long exposures (e.g. a long 10-min exposure, or the average of
several short exposures), thus improving the precision of 
the observations \citep{dum11}. We verified this with our Keck
data taken during 4 nights in January 2012. 
Each night we took three consecutive short ($\sim$3 min) exposures, 
so that the average exposure for a given night has a much lower contribution 
from stellar noise (lower panel of Fig.~\ref{radvelmcd}). 
The night-to-night scatter is only 3.3 m s$^{-1}$, in reasonable agreement with 
our observational error bar of 2.8 m s$^{-1}$.
The Keck measurements taken in February 2011 and in January-February 2012 
clearly rule out hot Jupiters, and we can even eliminate 
the presence of planets with masses as low as Neptune in the inner 0.04 AU
(lower panel of Fig. ~\ref{radvelmcd}).

Considering that velocity semi-amplitudes about two-three times the 
typical measurement precision (in our case $\sim$4 m s$^{-1}$) can be
detected with confidence, we can rule out the presence of
a stellar companion and of nearby giant planets. As shown in Fig.~\ref{planetsmcd},
where the 2-$\sigma$ (dashed line) and 3-$\sigma$ (solid line) sensitivity of 
our observations is shown 
(computed as the planetary mass that would introduce detectable radial velocity variations),
there is no indication for a Jupiter-mass planet in the terrestrial planet
region ($<$3 A.U.), and even a less massive giant planet such as Saturn may be ruled out
(at the 2-$\sigma$ level) inside 1 A.U. So, the inner region around HIP 56948 seems free from giant planets.
Examples of some circular orbits for a Jupiter-mass planet at 1 AU (Earth's distance
from the Sun), 1.5 AU (Mars' distance from the Sun)
and 3 AU, are shown in the upper panel of Fig.~\ref{radvelmcd}. We have verified for a range of
planetary masses, distances, random observing times and including error bars in the measurements,
that no giant planet is present, although notice that since our data are sparsely sampled, 
some giant planets (especially on orbits with $e > 0$) may have escaped detection.
Once more radial velocity data are obtained in the coming years
we intend to perform detailed simulations to see which kind of
planets we could have missed. With the existing dataset, we see
no evidence of giant planets in the inner region around HIP 56948.

Providing we maintain a radial velocity precision of about 4-5 m s$^{-1}$ for our observations, 
in a decade or so we should be able to detect (or rule out) 
the presence of a Jupiter twin, i.e., a Jupiter-mass planet 
orbiting at 5 A.U. from HIP 56948.

We will discuss the implications of the constraints we have deduced 
which limit the presence of massive planets in the inner regions around HIP 56948
 further in Sect. 4, in combination with 
our findings of the chemical similarities between HIP 56948 and the Sun.

\section{Abundance analysis}

The abundance analysis is based on the Keck HIRES spectra, following
our previous differential work on solar twins
\citep{mel06,mel07,mel09,ram09}, where the solar and stellar spectra
are measured in exactly the same way. 

An initial set of equivalent width (EW) measurements was obtained by
fitting gaussian profiles with the automatic routine ARES \citep{sou07}. 
We computed the relative difference 
in equivalent width between HIP 56948 and the Sun, $\delta W_r =
(EW^* - EW^\odot)/EW^\odot$, and lines
with $\delta W_r$ deviating from the median $<\delta W_r>$ for a given species,
were measured by hand both in HIP 56948 and the Sun. 
All weak lines (EW $<$ 10 m\AA) and the lines of
species with only a few lines available, were also measured by hand.
About 20\% of the EW measurements needed to be checked according to
our initial empirical analysis, although note that this procedure
only reveals the most obvious outliers. 
The deviating automatic ARES measurements could be due to any of these causes: 
{\it i)} the number of components found by ARES in the local fitting window is not exactly the 
same in the spectrum of HIP 56948 and the Sun, 
{\it ii)} the 2nd-order polynomial used to fit the local continuum could be somewhat different in both stars, 
{\it iii)} contamination by telluric lines do not fall exactly on the same place in both stars. 
Indeed, most faulty automatic measurements occur either in the blue, 
where the spectrum is more crowded, or in the red, where telluric contamination is higher.

An initial model atmosphere analysis is performed using the
improved EW measurements and again we look for outliers
from the mean abundance for each species, and they are 
checked by hand. In most cases the revised measurements result
in a reduced scatter. The improved EW is mainly due to more
consistent manual measurements, which are performed using exactly the 
same continuum for the Sun and the twin, and also the same part of the 
profile, or the same treatment of blends. This is not always the 
case with the automatic ARES measurements
The final analysis is performed using the
set of revised equivalent widths, except for lithium which
is analyzed using spectral synthesis.
The adopted EW for HIP 56948 and the Sun are given in Table \ref{tabew}.

We use 1D Kurucz overshooting model atmospheres \citep{cas97},
as well as MAFAGS-OS \citep{2004A&A...420..289G} and MARCS
\citep{gus08} 1D LTE models. 
The models have different mixing-length approaches,
as explained in detail in the above references.
The differences in stellar parameters, 
between HIP 56948 and the Sun, are small,
therefore essentially the same results
(within $\sim$0.001 dex) are obtained with
either Kurucz or MAFAGS-OS models.

The analysis has been performed both in LTE and NLTE.
For the LTE calculations we used the 2002 version of MOOG
\citep{sne73}, while the NLTE calculations are described in Sect. 3.3.

\subsection{Stellar parameters}

The stellar parameters adopted for the Sun are \teff = 5777 K and
log $g$ = 4.44 \citep[e.g.][]{cox00}. With \teff and log $g$
set, the microturbulence velocity (v$_t$) is found by requiring no 
dependence of $A$(Fe)\footnote{$A$(X) $\equiv$ log($N_{\rm X}$/$N_{\rm H}$) + 12} with 
reduced equivalent width EW$_r$ (= EW / $\lambda$) for FeI lines.
We found v$_t^\odot$ = 0.99 km s$^{-1}$ both for Kurucz and MAFAGS models.
The above set of parameters (5777 K, 4.44 dex, 0.99 km s$^{-1}$)
yielded the zero-point solar
abundances for each line $i$, $A_i^\odot$ for 
a given model atmosphere.

For HIP 56948, an initial set of stellar parameters was found in LTE using Kurucz models.
The {\it relative} spectroscopic equilibrium was achieved using differential abundances $\delta A_i$ for 
each line $i$, 

\begin{equation}
\delta A_i = A_i^* - A_i^\odot.
\end{equation}

Thus, the effective temperature is found by imposing the 
relative excitation equilibrium of $\delta A_i$ for FeI lines:

\begin{equation}
 d(\delta A_i^{\rm FeI}) / d(\chi_{\rm exc}) = 0 ,
\end{equation}

\noindent while the surface gravity (log $g$) is obtained 
using the relative ionization equilibrium. 
Usually, this is done using FeI and FeII, but in our case we verified that within the 
error bars the ionization balance is fulfilled simultaneously for Fe, Ti and Cr. Therefore, we use
the mean relative ionization equilibrium between FeI and FeII, TiI and TiII, and CrI and CrII:


\begin{eqnarray}
{\Delta^{\rm FeII-FeI} \equiv   < \delta A_i^{\rm FeII}> - < \delta A_i^{\rm FeI} > }  \nonumber \\
{\Delta^{\rm TiII-TiI} \equiv   < \delta A_i^{\rm TiII}> - < \delta A_i^{\rm TiI} > }  \nonumber \\
{\Delta^{\rm CrII-CrI} \equiv   < \delta A_i^{\rm CrII}> - < \delta A_i^{\rm CrI} > }  \nonumber \\
\Delta^{\rm II-I}  \equiv ( 3\Delta^{\rm FeII-FeI} \; + \; 2 \Delta^{\rm TiII-TiI} \; + \; \Delta^{\rm CrII-CrI}) / 6 \; = \;  0,
\end{eqnarray}

\noindent with the weights arbitrarily chosen, but reflecting increasingly larger errors for
FeI/FeII, TiI/TiII and CrI/CrII.

The microturbulence velocity v$_t$ was obtained when the differential abundances
$\delta A_i^{\rm FeI}$ show no dependence with reduced equivalent width EW$_r$:

\begin{equation}
 d(\delta A_i^{\rm FeI}) / d(EW_r) = 0 .
\end{equation}

The spectroscopic solution is found when the three conditions 
above (eqs. 2-4) are satisfied simultaneously, and when the metallicity
obtained from the iron lines is the same as that of the 
input model atmosphere. 
Notice from the equations above that our work is strictly differential, 
i.e., we do not enforce absolute spectroscopic equilibrium,
which may be difficult to achieve (both in LTE and NLTE)
even in the Sun \citep{mas11,ber12}.

The initial LTE solution with Kurucz models showed that indeed HIP 56948 is 
extremely similar to the Sun, with differences (HIP 56948 $-$ Sun)
in \tsin/log$g$/[Fe/H]/v$_t$ of only 17 K / 0.02 dex / 0.02 dex / 0.01 km s$^{-1}$.
We also tried the MAFAGS-OS models and the same stellar parameters 
were obtained, with a negligible difference of $\pm$0.001 dex in 
[Fe/H] from FeI and FeII lines with respect to the Kurucz models,
and with a difference in the spectroscopic \teff of only 0.1 K.

\begin{figure}
\resizebox{\hsize}{!}{\includegraphics{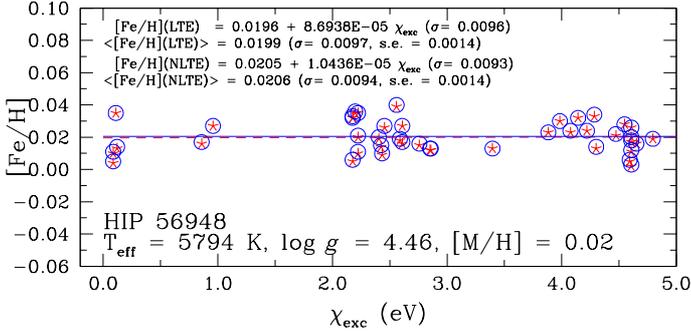}}
\caption{
Iron abundances versus excitation potential of FeI lines
in LTE (red stars) and NLTE (blue circles). The blue solid and
the red dashed lines show the fit in NLTE and LTE, respectively.
The line-to-line scatter in NLTE is only $\sigma$ = 0.009 dex
and the standard error is 0.001 dex.
}
\label{texc}
\end{figure}

Given the very similar stellar parameters between HIP 56948 and the Sun, 
we do not anticipate NLTE considerations to produce significant changes to the 
stellar parameters. Indeed, the NLTE Fe abundances indicate the same 
\teff within 0.4 K,
with the LTE temperature being slightly cooler. In Fig.~\ref{texc}
we show the individual iron abundances as a function of excitation potential of the FeI lines, 
both in LTE and NLTE. The NLTE abundances result in ever so
slightly lower line-to-line scatter ($\sigma$ = 0.009 dex, s.e. = 0.001 dex) than in LTE.

Unfortunately the trigonometric Hipparcos parallax for HIP 56948 is not
known with enough precision  to provide better constraints than our spectroscopic value.
The Hipparcos value is log $g$ = 4.37$\pm$0.07 (Table \ref{tabparam}),
which agrees within 1-$\sigma$ with our result from spectroscopy (log $g$ = 4.46$\pm$0.02).
The above error in the trigonometric log $g$ is due to both the uncertainty in 
the Hipparcos parallax and typical errors in photometric temperatures \citep[e.g.][]{mel10b}. Adopting instead
our more precise stellar parameters we obtain a Hipparcos-based gravity of log $g$ = 4.37$\pm$0.05, 
which agrees with our spectroscopic gravity within 1.3-$\sigma$.

Notice that the Hipparcos parallax (15.68$\pm$0.67 mas, according to the
new data reduction by \citealp{van07}) implies
a distance of 63.8$\pm$2.7 pc for HIP56948. We can obtain an independent
estimate of this distance assuming that the absolute magnitude of
HIP56948 is identical to solar; i.e. M$_{\rm V}$=4.81 \citep{bes98}.
Since V=8.671$\pm$0.004 and the error
in the absolute magnitude from the uncertainty in our stellar parameters is 0.055 mag
(see eq. 8), we derive a distance of 59.2$\pm$1.5 pc. Thus, there is
agreement within 1.1$\sigma$ for both distance estimates. 

The adopted stellar parameters, errors, and comparison with other estimates, are given in Table \ref{tabparam}.
Overall there is a good agreement (within the error bars) with other independent estimates.
The errors depend on the quality of the spectra of both HIP 56948
and the Sun. Since the S/N is very high ($\gtrsim$600) we can put
stringent constraints on the stellar parameters. Nevertheless, we
are also limited by the degeneracy between \tsin, log $g$, [Fe/H] 
and v$_t$, which increases the errors. 
As described in the appendix C, for a fixed log $g$ (or a small range of log $g$ values), 
we could determine \teff to within 0.8 K, 
while if \teff is keep fixed (or within a small range), log $g$ could be determined 
to within 0.006 dex. Regarding the microturbulence, for a fixed \teff and log $g$, 
$v_t$ could be determined to within 0.0004 km s$^{-1}$. Due to the degeneracies 
between stellar parameters and to the observational uncertainties, actually the errors 
are considerably larger (Table \ref{tabparam}), 7 K, 0.02 dex and 0.01 km s$^{-1}$ for \tsin, log $g$ and $v_t$, respectively. 
A detailed description of the determination of stellar parameters 
and the uniqueness of our solution, is given in the appendix C.

As a further check of the effective temperature, we have tried to compare
synthetic H$\alpha$ profiles, computed using MAFAGS-OS models, 
to the observed H$\alpha$ profiles in HIP 56948 and the Sun.
Unfortunately H$\alpha$ falls on an order too close to the edge
of the chip, making it difficult to normalize that region properly.
Despite the above problem, our tests using H$\alpha$ indicate
that indeed HIP 56948 is somewhat hotter ($\sim$ +20$\pm$20 K) than the Sun,
in agreement with what is found from iron lines.

\subsection{LTE abundances}

The adopted atomic data is presented in Table \ref{tabew}. Whenever
possible we use laboratory oscillator strengths, or theoretical
$gf$-values normalized to laboratory data \citep[e.g.,][]{fw06,mb09}. However, the input 
$gf$-values are not critical, as they cancel out in
the line-by-line differential abundances $\delta A_i$.
The interaction constants $C_6$ were computed from the broadening
cross-sections calculated by \cite{bar00} and \cite{bar05},
using the transformation given in \cite{mb09}. If broadening cross-sections 
were not available, we multiply the classical \"Unsold constant by 2.8.
The adopted $C_6$ values are given in Table \ref{tabew}.

The mean $<\delta A_i>$ and the standard deviation are computed
for each atomic species, so that we readily identify any outliers.
The suspicious measurements are checked by hand, both in the
Sun and HIP 56948, using insofar as possible the same
measurement criteria, i.e., the same continuum regions are
adopted and exactly the same part of the line profile is used
for the gaussian fit.

Equivalent widths were used to obtain abundances for all elements 
except for lithium, which was analyzed using spectrum synthesis,
as in our previous work on solar twins \citep{mel06,mel07,bau10}.
The line list used for spectral synthesis is presented
in Table~\ref{listli}. The data for the Li doublet was taken
from the laboratory data presented in \cite{and84}.
Although \cite{smi98} and \cite{hob99} reported new $gf$-values based on 
theoretical calculations, the difference with our adopted laboratory
values is only $\sim$1\%.
Other atomic lines near the Li feature were taken from
\cite{man04} and the Kurucz\footnote{http://kurucz.harvard.edu/}
and VALD \citep{kup00} databases, adjusting in some cases their $gf$-values 
to better reproduce the solar spectrum.
Molecular lines of CN \citep{mb99,man04} and C$_2$ \citep{mc07,ma08} 
were also included in the spectral synthesis.

The LTE abundances were computed using Kurucz models and
checked using MAFAGS-OS models. 
The agreement is excellent, with a mean difference (MAFAGS $-$ Kurucz) of
only $-$0.0010 dex in the differential abundances, and with a
element-to-element scatter in the differential abundances of only 0.00075 dex
(0.17\%), which is quite remarkable and shows the weak dependence
of our strict differential analysis to the adopted model atmosphere.
\footnote{The MARCS model and the mean atmospheric structure of a 
3D model atmosphere \citep{asp09}, are much closer to the MAFAGS-OS model
than to the Kurucz overshooting model, thus the effects 
of using either the MARCS or the 3D model would be even smaller than for the 
comparison between MAFAGS-OS and Kurucz overshooting models.}

The LTE differential abundances are provided in Table \ref{abund},
but note that the adopted abundances are those based on NLTE
(when available). The observational errors, which depend mainly
of the quality (S/N) of the spectra of both HIP 56948 and the Sun,
are given also in Table \ref{abund}. The observational error is adopted as 
the standard error (= $\sigma/\sqrt n$) when more than three lines of a given
species are available. Otherwise we assumed a minimum value of
$\sigma_{\min}$ = 0.009 dex (s.e. = 0.005 dex for 3 lines; s.e. = 0.006 dex for 2 lines),
which is the typical line-to-line scatter for species with more than 
10 lines available. When only 2 or 3 lines were available we adopted the 
maximum value of the observed $\sigma$ and $\sigma_{\min}$, i.e., $max$($\sigma$, 0.009 dex).
If only one line was available, we estimated
the error by performing a number of measurements with different assumptions
for continuum placement (within the noise of the spectra) and profile fitting.

In addition to the observational errors, we also present the errors
due to uncertainties in the stellar parameters in Table \ref{abund},
where the total error is also given. 
As shown in appendix B, our very small observational errors ($\sim$0.005 dex) are plausible.

\subsection{NLTE abundances}

For most of the elements we have been able to account for departures from local 
thermodynamic equilibrium (NLTE). For Li, C, O and Na we have employed 
MARCS model atmospheres \citep{gus08} and the model atoms described in 
\cite{lin09}, \cite{fab06,fab09} and \cite{lin11a}, 
respectively, using up-to-date radiative and collisional data. We note that for 
Li and Na, quantum mechanical estimates of the cross-sections for collisions 
with both electrons and hydrogen are available.
For O we adopt a scaling factor 
S$_{\rm H}$ of 0.85 to the classical formula of \cite{drawin68,drawin69} for excitation and 
ionization due to inelastic H collisions as empirically determined from the 
solar center-to-limb variation by \cite{per09}. 
For C we adopt S$_{\rm H}$ = 0.1 in the absence of similar empirical evidence. 
For these elements 
the NLTE calculations were performed with the 1D statistical equilibrium 
code MULTI \citep{car86}.

In addition, we performed NLTE calculations for
Mg, Al, Ti, Cr, Mn, Fe, Co, and Ba,
using the revised version of the DETAIL
code \cite[originally published in][]{1981PhDT.......113G} and 
the SIU code (J. Reetz, unpublished). We
used MAFAGS-OS 1D LTE model atmospheres \citep{2004A&A...420..289G} provided by F.
Grupp (private communication).
The differences in abundances required to equalize NLTE and LTE equivalent widths
are defined as NLTE corrections. 

For Mg and Al, the model atoms were kindly provided by T. Gehren; 
those models were previously used in the spectroscopic analysis of solar-type stars given in
\citet{2004A&A...413.1045G, 2006A&A...451.1065G}. Atomic models for Cr, Mn,
Co, and Ti, were taken from \citet{2010A&A...522A...9B}, \citet{2008A&A...492..823B},
\citet{2010MNRAS.401.1334B}, and \citet{ber11}, respectively. To compute NLTE corrections for the
lines of Fe I/II, and Ba II, we constructed the model atoms from the
laboratory and theoretical data given in
NIST\footnote{http://www.nist.gov/physlab/data/asd.cfm} and
Kurucz\footnote{http://kurucz.harvard.edu/} databases.
For Fe I, we also used highly-excited predicted levels and transitions as recommended 
by \citet{2010IAUS..265..197M}, who showed that the inclusion of these data is necessary 
for a realistic representation of statistical equilibrium of Fe in the atmospheres of cool stars. 
Photoionization cross-sections for 
Fe I levels were taken from \citet{1997A&AS..122..167B}. 
The detailed description of the Fe model is given in \cite{ber12}.

A critical parameter in the statistical equilibrium calculations is the
efficiency of inelastic collisions with H I. In the absence of
quantum-mechanical data, we computed the cross-sections for excitation and
ionization by H I atoms from the formulae of \cite{drawin69}. Following
the above-mentioned studies, we adopted individual scaling factors S$_{\rm H}$ to
the Drawin-type cross-sections for each element: 
Mg (0.05), Al (0.002), Ti (0.05), Cr (0), Mn (0.05), Co (0.05). These scaling factors were determined by
requiring consistent ionization-excitation equilibria of the elements under
restriction of different stellar parameters. For Fe and Ba, we used S$_{\rm
H}$ = $0.1$, and $0.05$, respectively.

We performed another set of independent non-LTE calculations for Na
using MAFAGS-OS models and the model atom described by \cite{2006A&A...451.1065G}, but
the mean Na abundance only changes by 0.001 dex with respect to the value obtained 
using the most up-to-date model atom by \cite{lin11a}.

Besides the elements above (Li, C, O, 
Na, Mg, Al, Ti, Cr, Mn, Fe, Co, and Ba) 
for which specific NLTE calculations were performed for the present work, 
we estimate NLTE corrections for K, Ca, Zn and Zr, 
using the grid of NLTE corrections computed 
by \cite{tak02}, \cite{mas07}, \cite{tak05}
and \cite{vel10}, respectively. Nevertheless, the differential NLTE 
abundance corrections are negligible for these elements ($\le$0.001 dex).

All the differential NLTE corrections (HIP 56948 - Sun) are given in Table \ref{tabew},
except for Li. For this element the differential NLTE correction amounts to only -0.001 dex
and the adopted differential NLTE Li abundance is given in Table \ref{tabparam}.

\begin{figure}
\resizebox{\hsize}{!}{\includegraphics{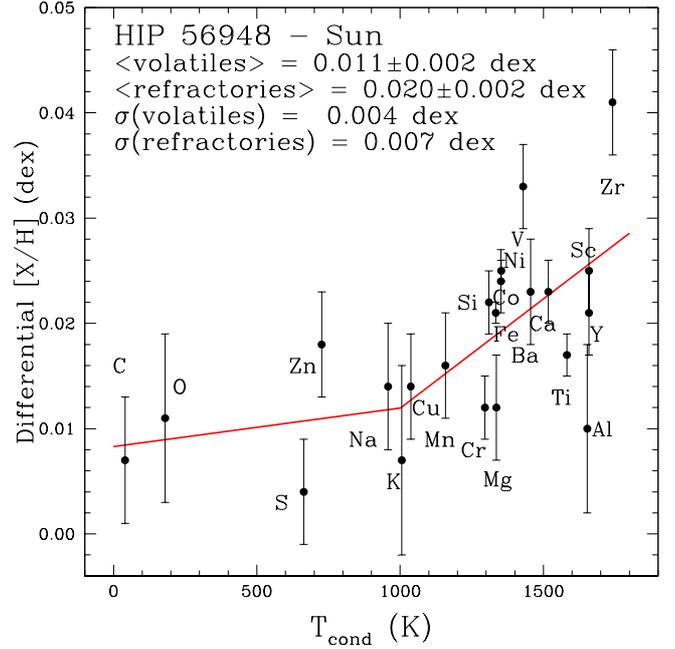}}
\caption{
Abundance pattern of HIP 56948 (circles) versus condensation temperature.
The solid line represents the mean abundance pattern. 
The error bars are based only on the observational uncertainties, which are $\sim$0.005 dex.
The low element-to-element scatter from the fit for the volatile
($\sigma$ = 0.004 dex) and refractory ($\sigma$ = 0.007 dex) elements
confirms the high precision of our work.
}
\label{hip56948a}
\end{figure}

\begin{figure}
\resizebox{\hsize}{!}{\includegraphics{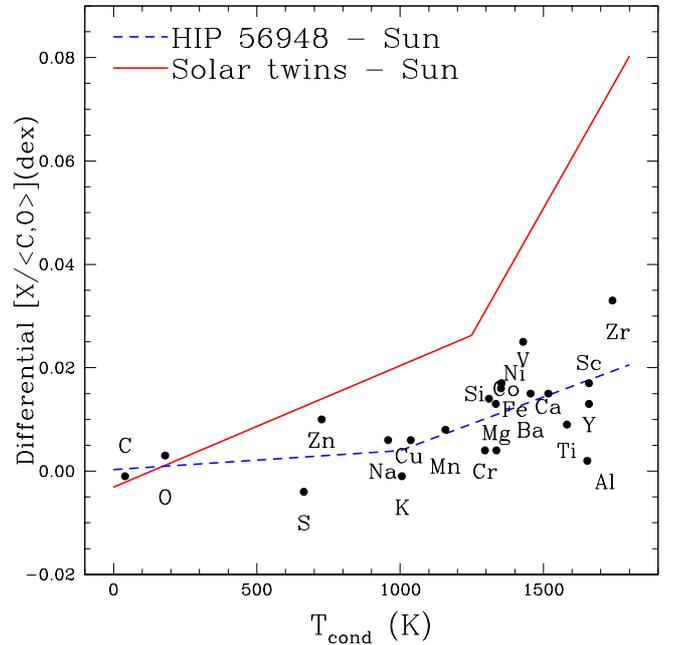}}
\caption{
Abundance pattern of HIP 56948 (circles) versus condensation temperature.
The average of the highly volatile (low T$_{\rm cond}$) elements C and O is used as reference.
The solid line represents the mean abundance pattern
of the 11 solar twins studied by \cite{mel09} and
the dashed line the fit to the abundance pattern of HIP 56948.
Clearly, HIP 56948 is much closer to the Sun than to 
other solar twins.
}
\label{hip56948tcond}
\end{figure}

\subsection{The abundance pattern of HIP 56948}

In Fig.~\ref{hip56948a} we plot the differential abundances [X/H] between HIP 56948 and the Sun
(circles) as a function of equilibrium condensation temperature \citep[T$_{\rm cond}$,][]{lod03}. 
The fit of [X/H] vs. T$_{\rm cond}$ is shown for volatile
(T$_{\rm cond} <$ 1000 K) and refractory (T$_{\rm cond} >$ 1000 K) elements.
\footnote{Notice that for the abundance pattern of HIP 56948 the best fit is found for a break at 
T$_{\rm cond}$ = 1000 K, while for the average of solar twins the break is at T$_{\rm cond}$ = 1200 K.
Although the break point was determined by fitting independently the refractory
and volatile elements, and choosing the break by minimizing the scatter
in both sides, with minor adjustments to provide the best match at the break point, 
our results are confirmed by a global fitting that assumes two linear
functions to fit the whole dataset by assuming that those functions are equal at the break point.}
As can be seen, the element-to-element scatter around the fit is
extremely small, only 0.004 dex for the volatiles and 0.007 dex for
the refractories. Both are of the same order as the observational
error bars, which are $\sigma_m$(obs) $\sim$ 0.006 dex for volatiles
and $\sigma_m$(obs) $\sim$ 0.004 dex for refractories,
hence showing that it is possible to achieve abundances with errors
as low as $\sim$0.005 dex.
If we take into account the errors due to uncertainties in the stellar
parameters (Table \ref{abund}), then the expected total median errors 
(including observational errors) are 0.007 dex for volatiles and 0.008 dex 
for refractories, i.e., somewhat higher than the observed scatter (0.004 and
0.007 dex, respectively) around the mean trends. That means that 
we may be slightly overestimating our errors, and that
a more representative total error for our abundances is $\sim$0.006 dex. 
As can be seen in Table \ref{abund}, it is possible
to obtain a total error as small as 0.004 dex (Si) or 0.005 dex (Ni), and for
several elements errors as small as 0.006 dex can be achieved (Ca, Cr, Fe, Co, Zn).

The mean [X/H] ratio of the volatile and refractory elements is 
$<$[volatiles/H]$>$ = 0.011 (s.e. = 0.002) and 
$<$[refractories/H]$>$ = 0.020 (s.e. = 0.002), respectively.
Thus, the mean difference between refractories and volatiles in HIP 56948
amounts to only 0.009 dex, i.e., HIP 56948 has an abundance
pattern very similar to solar.

In Fig.~\ref{hip56948tcond} we plot the differential abundances [X/$<$C,O$>$] between HIP 56948 and the Sun
(circles) as a function of condensation temperature. 
Here the average of carbon and oxygen, $<$C,O$>$, is chosen as the
reference, as those elements should not be depleted in the solar atmosphere \citep{mel09}.
We also show with a solid line the mean\footnote{the robust estimator {\em trimean} is used} 
abundance pattern of the 11 solar twins studied by \cite{mel09}.
The dashed line represents the mean behavior of HIP 56948
shown in Fig.~\ref{hip56948a}. This figure shows clearly that the abundance
pattern of HIP 56948 is much closer to the Sun than to the 
mean abundance pattern of other solar twins.

The volatile elements with T$_{\rm cond} <$ 1000 K have 
a similar behavior in HIP 56948 and the Sun, while the refractory elements are
depleted (with respect to other solar twins) both in HIP 56948 and the Sun, although they are
somewhat less depleted (by $\sim$0.01 dex) in HIP 56948.
Therefore, it seems that 
somewhat less dust was formed around HIP 56948 than around the Sun.
Interestingly, in the sample of solar twins studied so far, 
the Sun seems to be the most depleted in refractories.
If this peculiar abundance pattern is related to the formation of rocky planets
\citep{mel09,ram09,ram10,cha10}, that signature
could be used to search for possible candidates to host terrestrial planets.
Below, we discuss in detail various other possibilities that could
cause abundance anomalies.

\section{Discussion}

From the determination of isotopic abundances in
meteorites it has been shown that short-lived radionuclides
were present in the early stages of the solar system, 
perhaps due to pollution by an asymptotic giant branch (AGB) star 
\citep{was94,bus99,tri09} or by supernova \citep{ct77,bf98,oue10},
although other causes may be possible.
Thus, before discussing the terrestrial planet formation scenario,
we assess if the solar anomalies may be due to other causes
such as pollution by intermediate or high mass stars.

\begin{figure}
\resizebox{\hsize}{!}{\includegraphics{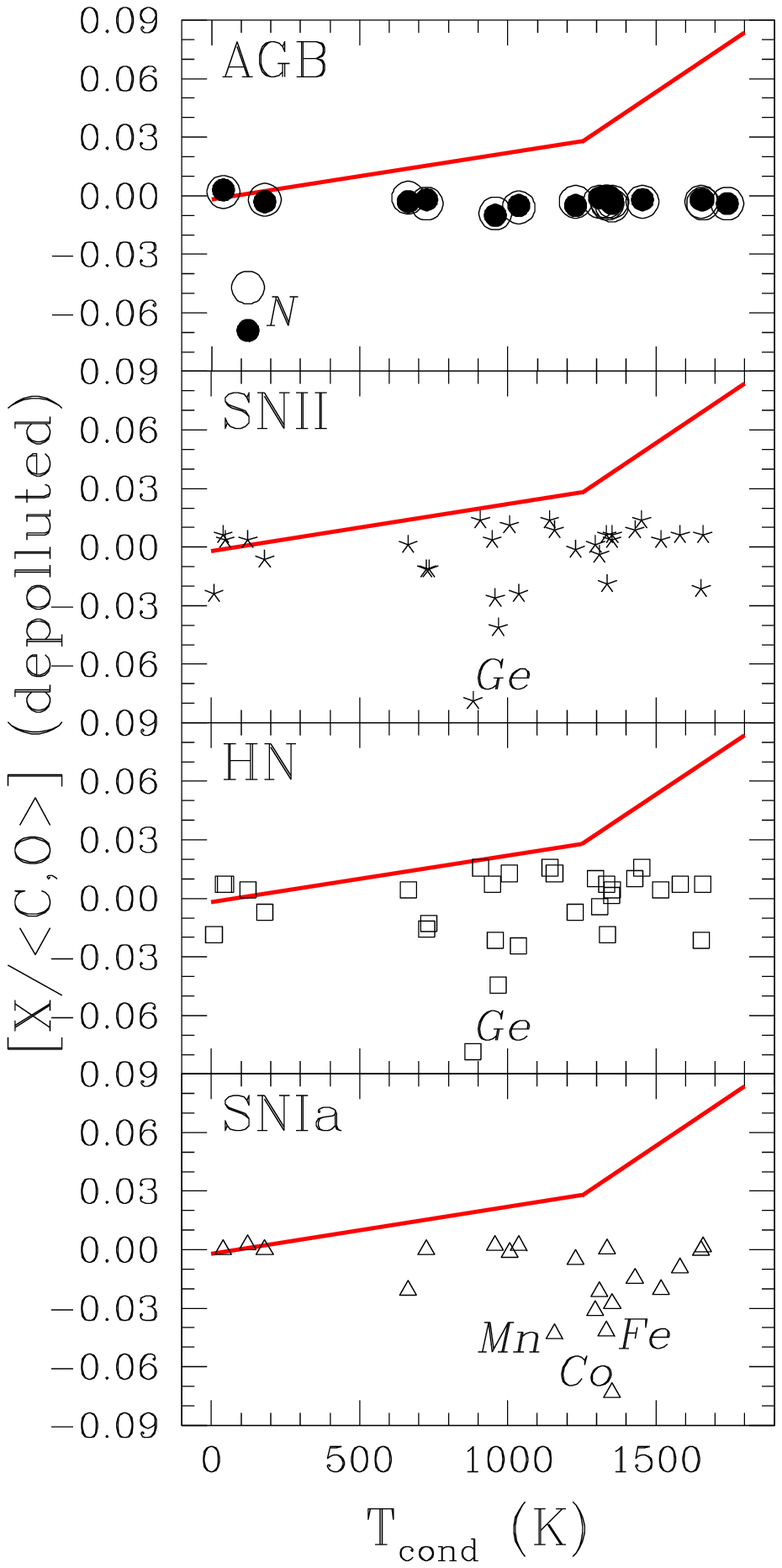}}
\caption{
Abundance ratios obtained after de-polluting the solar nebula from
contamination by an AGB star (circles), 
SNII (stars), hypernova (squares) and SNIa (triangles).
In the top panel it is shown the effect of
adopting different solar abundances 
(open circles: \cite{and89}; filled circles: \cite{asp09}).
The solid line represents the mean abundance pattern
of 11 solar twins relative to the Sun \citep{mel09}. None of the
pollution scenarios can explain the trend with condensation temperature.
For clarity, the abundance ratios of SNII, HN and SNIa
have been divided by 4, 3.5 and 14, respectively.
The chemical elements that change the most are labeled.
}
\label{pollution}
\end{figure}

\subsection{Abundance anomalies: AGB/SN pollution or terrestrial planets?}

In \cite{mel09} we argued against the hypothesis that the abundance anomalies found 
in the Sun could be due to galactic chemical evolution effects
or supernova pollution, as the solar chemical peculiarities 
do not seem to follow those abundance patterns. Here, we 
study in more detail whether the anomalies could be due
to pollution by an AGB star, thermo-nuclear supernovae (SNIa), 
core-collapse supernovae (SNII) or a hypernova (HN),
by subtracting the yields of those objects to the solar nebula
(i.e., ``de-polluting'' the solar abundances)
and comparing the results to the pattern of the solar twins.

Following the tentative AGB pollution scenario of \cite{tri09}, 
we use a dilution factor of 1 part of AGB material
per 300 parts of original solar nebula material (equivalent
to mixing 0.0185 M$_\odot$ of AGB ejecta), for which 
solar abundances from \cite{and89} were adopted. 
Note that the use of the new solar abundances by \cite{asp09} do not
have a significant impact on the abundance trend, as shown in
Fig.~\ref{pollution} (top panel). 
For SNIa, SNII, and HN we assume that the same amount of
mass as used in the AGB scenario is mixed into 1M$_\odot$ of
solar system material. For consistency with the AGB scenario
we use the solar abundances from \cite{and89}.

The amount of material that would actually be injected 
is uncertain. 
There is no reason that a SN would inject about the same 
amount of mass as an AGB star. \cite{young11} note that
SN are more likely to pollute the entire molecular star 
forming cloud, not an individual protosolar nebula. They
estimate that a $\sim 1$\% enrichment of the protosolar molecular 
cloud by ejecta from SNII can account for the
oxygen isotope ratios measured in the solar system.
However as we are only interested in the qualitative impact
(trend with condensation temperature) that such yields have 
on a protosolar nebula and not the quantitative
details, the actual amount of pollution from an individual SN or
AGB model should not strongly impact our conclusions. 
The same can be said for the exact amount of the AGB and SN yields,
which are inherently quite uncertain.

A massive AGB star of 6.5 M$_\odot$ was chosen, as it has a short
lifetime ($\sim$55 Myr). Our AGB model and yields are described in
\citet{kar07}, \citet{tri09} and \citet{kar10}.
The results after removing the AGB pollution are
shown in Fig.~\ref{pollution} (top panel). As can be seen, contamination by a massive AGB
star cannot explain the abundance trend with T$_{\rm cond}$. 
The main signature from an hypothetical 6.5-M$_\odot$ AGB star 
would be a large change in nitrogen.

We have also considered pollution by SNII, HN, and SNIa, 
according to the yields described in \citet{kob06}, \citet{kob09} and
\citet{kob11}.
\citet{mik07} suggest that a supernova of a massive star of
at least 20 M$_\odot$ was responsible for the anomalies 
of short-lived radionuclides discussed above,
so for the SNII and HN, a 25 M$_\odot$ was adopted.
As seen in Fig.~\ref{pollution}, the trend with condensation temperature cannot be
explained by either SNII or HN.

Finally, for the binary system leading to a SNIa,
we choose a system composed of a 3M$_\odot$ of primary star + 1.3M$_\odot$ of secondary star,
with a metallicity of 0.1 Z$_\odot$ for the progenitors, which are supposed to have been born 4.5Gyr ago.
The binary system evolves to $\sim$1M$_\odot$ C+O white dwarf plus 1.3 M$_\odot$ of secondary star,
then to 1.374 M$_\odot$ SNIa + 0.9 M$_\odot$ remnant of the secondary star.
Thus, for the contamination by ejecta of SNIa, we
include 1.374 M$_\odot$  of processed metals by SNIa plus
(3 + 1.3 $-$ 0.9 $-$ 1.374) M$_\odot$ of unprocessed matter 
in the stellar winds.

Again, pollution by SNIa cannot explain the peculiar solar abundance pattern (Fig.~\ref{pollution}).
We also tried other combinations for the progenitors of SNIa, but these yields
also do not reproduce the trend with condensation temperature shown in Fig.~\ref{pollution}.

Therefore, we conclude that the peculiar solar abundance pattern cannot be due
to contamination by AGB stars, SNIa, SNII or HN, as predicted by
state-of-the-art nucleosynthesis models.
We emphasize that our results do not rule out that a SN or AGB star
contaminated the proto-solar nebula to cause the observed
isotopic anomalies in meteorites, rather, we discard pollution as a viable
explanation of the peculiar elemental abundances in the Sun.

\subsection{The abundance pattern of HIP 56948 and terrestrial planets}

\begin{figure}
\resizebox{\hsize}{!}{\includegraphics{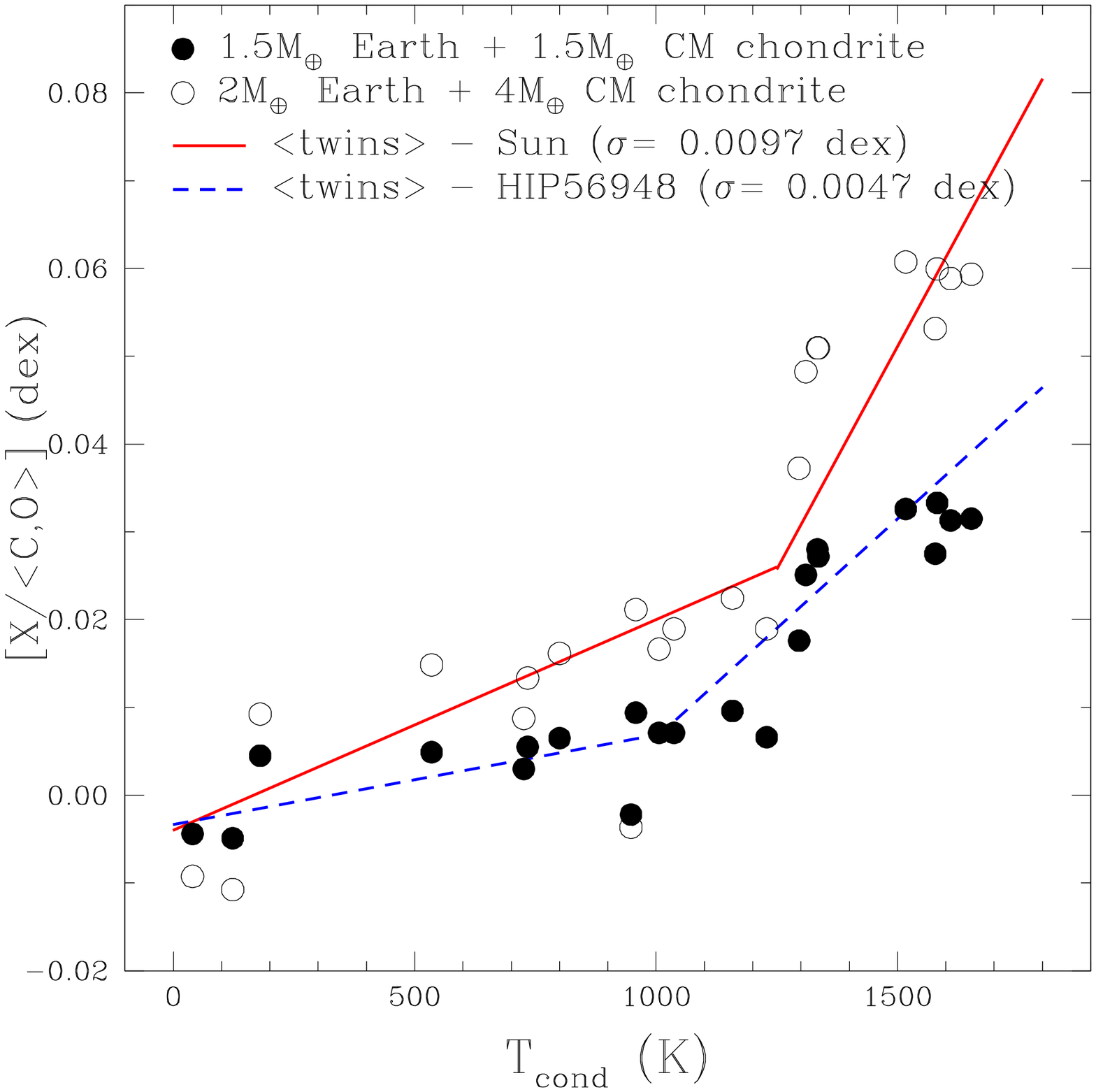}}
\caption{
Composition of the solar twins with respect to
HIP 56948 (dashed line) and the Sun (solid line).
The open circles show the effect of adding 6 M$_\oplus$ of 
a mix of Earth-like and meteoritic-like material 
\citep{cha10} to the convection zone of the present Sun 
\citep{asp09}, and the filled circles the effect of
adding 3 M$_\oplus$ of rocky material to the convection
zone of HIP 56948.
Abundances are normalized with respect to $<$C,O$>$. 
Both the Sun and HIP 56948 require $\sim$2 M$_\oplus$ of 
Earth-like material, while the Sun requires much larger
quantities of chondrite material. Notice the
very small ($<$0.01 dex) element-to-element scatter.
}
\label{terra}
\end{figure}

As discussed above, pollution from stellar ejecta cannot explain the anomalous solar abundance pattern.
A possible explanation for the peculiarities is the formation 
of terrestrial planets \citep{mel09,ram09,ram10,gus10,cha10}.
The same planet formation scenario may be applied to HIP 56948. 
To verify this, we obtained the abundance ratios of the 
11 solar twins of \cite{mel09} relative to HIP 56948,
again using the average of C and O as reference.
The resulting mean trend is represented by a dashed line in Fig.~\ref{terra}. 
The trend of solar twins relative to the Sun is shown by a solid line.

Thus, the same terrestrial planet formation scenario could
be applied for HIP 56948, except that the Sun has
$\sim$0.01 dex smaller refractory-to-volatile ratio than HIP 56948.
So, the Sun could have formed more rocky planets than
HIP 56948, or perhaps slightly more massive rocky
planets than in HIP 56948. In any case, the overall 
amount of rocky material may have been higher around the Sun
than around HIP 56948.

Recently, \cite{cha10} has shown that a mix of
4 Earth masses of Earth-like and meteoritic-like material,
provides an excellent element-to-element fit for
the solar abundance anomalies for about two dozen chemical elements.\footnote{Notice 
that although this indicates that a certain amount of both Earth-like and 
meteoritic-like material may have been removed from the Sun (in comparison 
with the solar twins), this does not imply necessarily that the removed 
material was employed to form the terrestrial planets and asteroids.}
Interestingly, the above mixture reproduces the anomalies
significantly better than a composition based either only on Earth material
or only on carbonaceous-chondrite meteorites.
Using his detailed abundance pattern of Earth and 
carbonaceous chondrites, we can check whether the same mixture 
could fit HIP 56948. Indeed, as shown in Fig.~\ref{terra}, 
a 3-Earth-mass mixture of 
Earth plus meteorites (filled circles) provides an excellent fit
to the abundance pattern of the solar twins relative
to HIP 56948 (dashed line). The element-to-element scatter
of the Earth/chondrite mix with respect to the fit (dashed line)
is only 0.005 dex. Notice that the chemical elements
included in Fig.~\ref{terra} are not necessarily the same
ones studied in HIP 56948 because for the Earth/meteoritic
mix we can include other elements such as fluorine and uranium
\citep{cha10}.

The abundance pattern of the solar twins with respect to
the Sun (Fig.~\ref{terra}, solid line), seems to require more than
the four Earth masses suggested by \cite{cha10}. Note that
we are normalizing to $<$C,O$>$, while \cite{cha10} normalizes
to iron, and that the mean solar abundance trend adopted by
\cite{cha10} may be slightly different than ours, 
hence some differences are expected.
The fit for the Sun ($\sigma$ = 0.010 dex) is not as good
as for HIP 56948 ($\sigma$ = 0.005 dex). In total 6 M$_\oplus$ of
rocky material is needed to fit the Sun, while 3 M$_\oplus$  is needed
to fit HIP 56948. So, perhaps about twice more
rocky material was formed around the Sun than around HIP 56948.

Interestingly, both the Sun and HIP 56948 seem to require about the same 
mass of Earth-like material (2 and 1.5 M$_\oplus$, respectively), 
but the amount of CM chondrite material seems much higher for the 
Sun (4 M$_\oplus$) than for HIP 56948 (1.5 M$_\oplus$). Although 4 Earth masses 
of meteoritic material may seem too high considering the current mass of the asteroid belt, 
various models predict that the belt has lost most of its
mass \cite[e.g.,][]{wei77,wet89,wet92,pet99,cw01,min10}. 
In particular, the simulations by \cite{cw01} show that a few Earth masses of 
material from the belt can be removed, either via collisions with the Sun or 
ejected from the solar system. In contrast, material in the terrestrial-planet 
region had a good chance of surviving \citep{cw01}. Thus, albeit a large mass 
of material may have formed in the asteroid belt, most of the initial mass has already been removed.

The 2 M$_\oplus$ of Earth-like material seen in the chemical composition of the Sun is 
comparable with the total mass of the terrestrial planets in the solar system, 
and only somewhat higher than the amount of Earth-like material around HIP 56948. 
As shown above, the different slopes of the abundances of volatile and 
refractory elements (with  condensation temperature) may be used to constraint the 
type of solid bodies that were originated as a result of planet formation using
the different abundance patterns of the Earth and meteorites \citep{cha10}. 
Thus, the careful analysis of stellar chemical compositions offer the thrilling 
prospect of determining which type of rocky objects were formed around stars.

In line with our above findings on the similarities between HIP56948 and the Sun, 
the radial velocity monitoring of HIP 56948 (Section 2.2) shows no indication
of inner ($<$ 3 A.U.) giant planets, as massive as Saturn or Jupiter. 
Thus, the inner region around HIP 56948 can potentially host
terrestrial planets. The remarkable chemical similarities
between HIP 56948 and the Sun, also suggest that 
HIP 56948 may be capable of hosting rocky planets.

Also, metal-rich solar analogs without close-in giant planets seem to have
an abundance pattern closer to solar than stars with detected
giant planets \citep{mel09}.\footnote{Notice that, as discussed 
in \cite{sch11} and \cite{ram10}, the interpretation of abundance trends 
for metal-rich stars is complicated by Galactic chemical evolution processes.}

Our findings open the truly fascinating possibility of identifying Earth-mass 
planets around other stars based on a careful high resolution spectroscopic analysis 
of stellar chemical compositions. 
Once the Kepler mission \citep[e.g.][]{bor10} announces the discovery of Earth-sized planets 
in the habitable zones of G-type dwarfs, our planet signature could be verified 
by high precision chemical abundance analyses of those stars.
Although they are relatively faint (confirmed Kepler planet-hosting stars 
have a mean magnitude of $K_{\rm p}$ = 13.8$\pm$1.5),
recently \cite{one11} have shown that it is possible to achieve high precision (0.03 dex) 
differential abundances even in stars with V$\approx$15. They analyzed the faint (V = 14.6) 
solar twin M67-1194 using VLT/FLAMES-UVES, showing that it has a remarkable chemical similarity to the Sun.

\subsection{The mass, age, luminosity and radius of HIP 56948}

\begin{figure}
\includegraphics[width=9.5cm]{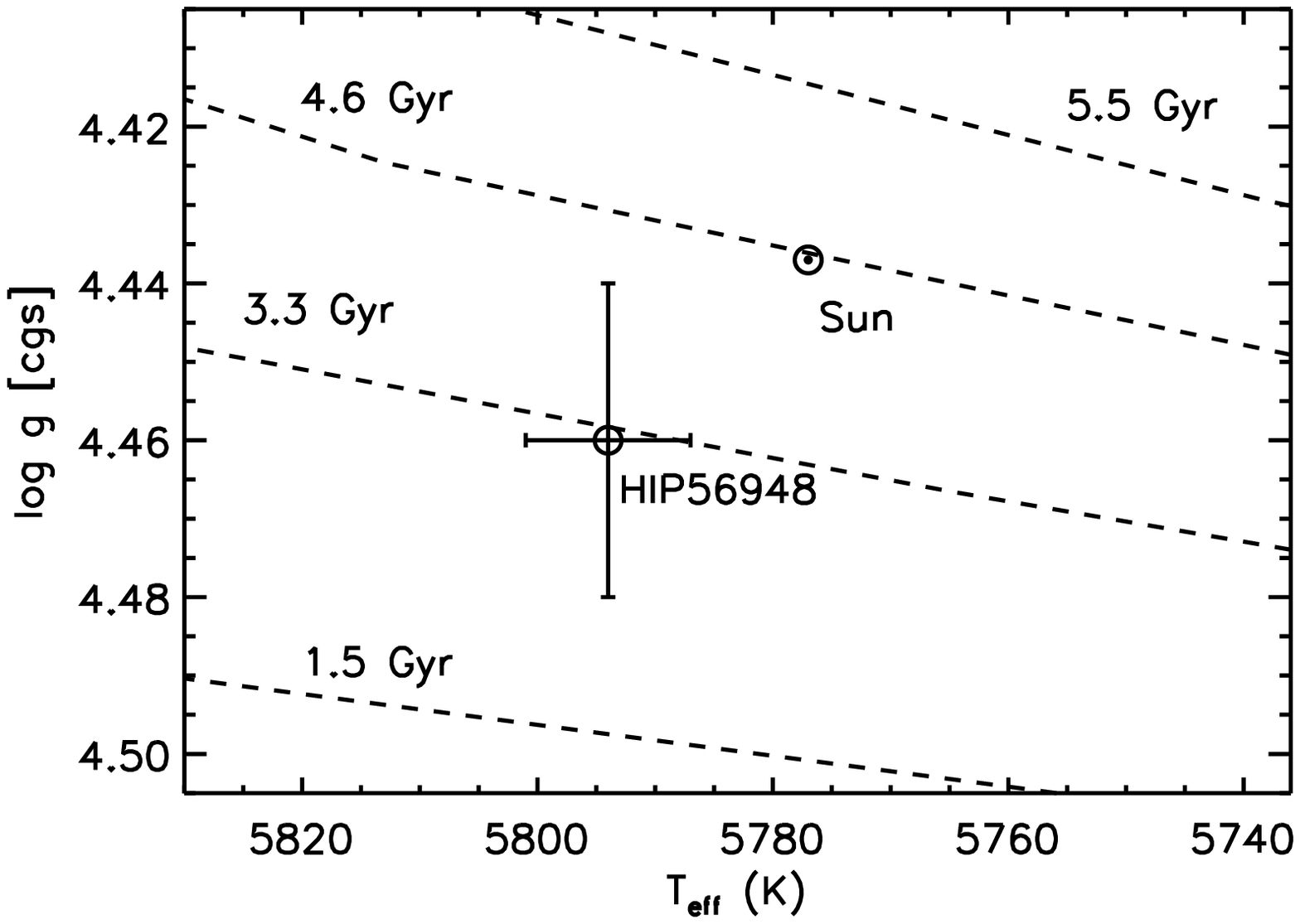}
\caption{Location of the Sun and HIP\,56948 on the HR diagram. Note the very small range of stellar parameters. 
Solar-metallicity isochrones of 1.5, 3.3, 4.6, and 5.5 Gyr are shown (dotted lines). 
The high precision of our derived stellar parameters for HIP\,56948 allows us to infer a reasonable estimate of its age from 
the theoretical isochrones, even though they are densely packed in this main-sequence region.}
\label{f:age}
\end{figure}

Contrary to commonly thought, reasonable estimates of the ages of main-sequence stars can be obtained 
using standard isochrone fitting techniques,\footnote{see \cite{bau10,mel10c,ben11,ram11,cha12}, for 
different applications of our isochrone ages.} provided the isochrones are accurate 
(i.e., calibrated to reproduce the solar age and mass) and the stellar parameters $\tef,\logg,\feh$ are known 
with extreme precision, as illustrated in Fig.~\ref{f:age}.

We used a fine grid of Yonsei-Yale isochrones \citep{yi01,kim02,dem04} 
with a step $\Delta$[Fe/H] = 0.01 dex around solar metallicity
($-0.15 \leq$ [Fe/H] $\leq +0.15)$ and a step of $\Delta$[Fe/H] = 0.02 dex
elsewhere. We adopt [$\alpha$/Fe] = 0 for [Fe/H] $\geq$ 0, 
[$\alpha$/Fe] = $-$0.3 $\times$[Fe/H] for $-$1 $<$ [Fe/H] $<$ 0,
and [$\alpha$/Fe] = +0.3 for [Fe/H] $\leq$ $-$1. 
The isochrones include a dependence between helium abundance $Y$ and metallicity $Z$
with a slope of 2, $Y$ = 0.23 + 2 $Z$ \citep{yi01}.

Our grid has been normalized to reproduce the solar 
age and mass from the input solar parameters (\teff = 5777 K, log $g$ = 4.44, [Fe/H] = 0.0). The normalization factor 
was found by performing small offsets around the solar \tsin, log $g$ and [Fe/H]
to see which offsets better reproduce the solar age and mass. 
We found that shifts in $\tef$ and log $g$ are not needed, as the best compromise
solution can be found with only a small shift of $-$0.04 dex in the observed [Fe/H],
meaning that the models are off by +0.04 dex in metallicity.
Thus, the input metallicity used to compare with the
isochrones is:

\begin{equation}
{\rm [Fe/H]^{input} = [Fe/H] - \; 0.04 \; dex.}
\label{eq:shift}
\end{equation}

That normalization gives a mean solar mass of 1.000 M$_\odot$ (to within 0.003 M$_\odot$)
and a mean solar age of $\sim$4.5 Gyr (to within 0.2 Gyr), which is in excellent
agreement with the age of the solar system \cite[$\sim$4.567 Gyr,][]{con08,ame10}. 
The calibration of the models to the solar mass
and age is valid for a broad range of errors of 
10-140 K in \teff and 0.01-0.10 dex both in log $g$ and 
[Fe/H]. Thus, after the zero-point shift of Eq.~\ref{eq:shift} is applied,
our resulting masses and ages are accurate.
If the input errors are much higher than those indicated above,
the zero point of the solutions need to be revised.\footnote{For
errors of 175 K in \teff and 0.1 dex both in log $g$ and 
[Fe/H], the solar mass and age would be about 0.99 M$_\odot$ and 4.8 Gyr,
while for an error of 250 K in \teff and 0.15 dex both in log $g$ and 
[Fe/H], the solar mass and age would be about 0.98 M$_\odot$ and 5.5 Gyr,
respectively. However, for typical errors of abundance analysis
of $\sigma$(\teff) $<$ 150 K, $\sigma$(log $g$) $\leq$ 0.1 dex and 
 $\sigma$([Fe/H]) $\leq$ 0.1 dex, no zero-point corrections (besides that
already used in Eq.~\ref{eq:shift}) would be needed to obtain accurate masses and ages.}

The isochrone points are characterized by effective temperature ($T$), logarithm of surface gravity ($G$), and metallicity ($M$),
with a step in metallicity of 0.01\,dex around [Fe/H] = 0.

We obtained an estimate of the age of HIP\,56948 from its isochrone age probability distribution (APD):
\begin{equation}
dP\mathrm{(age)} = \frac{1}{\Delta(\mathrm{age})}\sum_{\Delta\mathrm{(age)}} p\,(\tef,\logg,\feh,T,G,M)\ ,
\label{eq:dP}
\end{equation}
where $\tef, \logg, \feh$ are the observed stellar parameters, $\Delta\mathrm{(age)}$ is an adopted step in age from the grid of isochrones, and:
\begin{eqnarray}
p & \propto & \exp[-(\tef-T)^2/2(\Delta\tef)^2]\times   \nonumber \\
  &         & \exp[-(\logg-G)^2/2(\Delta\logg)^2]\times  \\
  &         & \exp[-(\feh-M)^2/2(\Delta\feh)^2]\ . \nonumber
\end{eqnarray}
The errors in observed stellar parameters are $\Delta\tef$, etc. 
The sum in Eq.~\ref{eq:dP} is made over a range of isochrone ages and in principle 
all values of $T,G,M$. In practice, however, the contribution to the sum from isochrone points 
farther away than $\tef\pm3\Delta\tef$, etc., is negligible. Therefore, the sum is limited to 
isochrone points within a radius of three times the errors around the observed stellar parameters. 
A similar formalism allows us to infer the stellar mass. The probability distributions are normalized 
so that $\sum dP=1$. The most probable age and mass are obtained from the peaks of 
these distributions while $1\,\sigma$ and $2\,\sigma$ Gaussian-like lower and upper limits 
can be derived from the shape of the probability distributions.

\begin{figure}
\includegraphics[width=9.5cm]{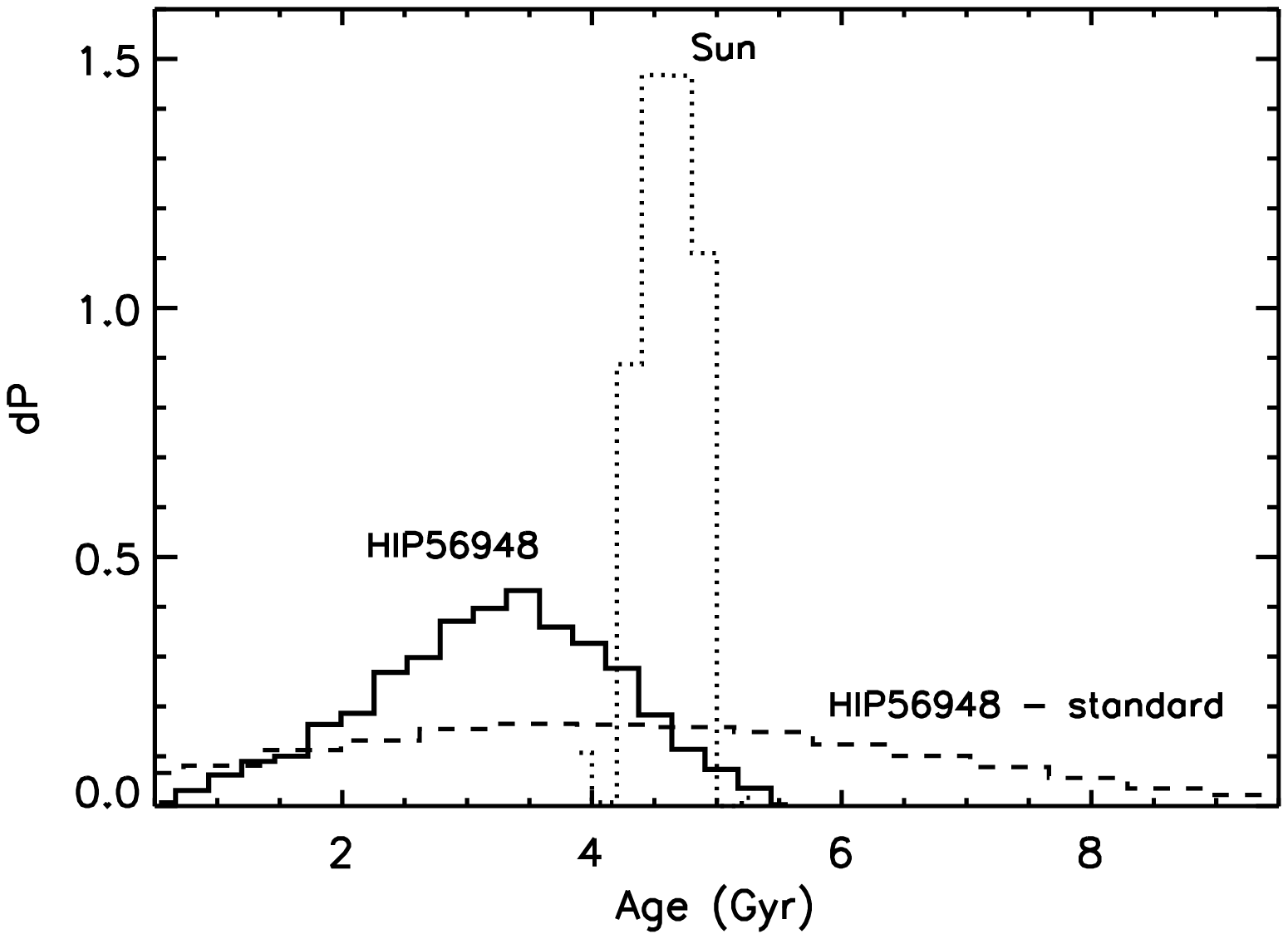}
\caption{Age probability distributions (APDs) for the Sun (dotted line) and HIP\,56948 (solid line). 
The dashed line corresponds to the APD that we would obtain for HIP\,56948 if we had derived 
its stellar parameters from photometric data and \textit{Hipparcos} parallax (the ``standard'' method) 
instead of performing our very precise strict differential analysis. Clearly, our differential
approach gives much better results.}
\label{f:agepdf}
\end{figure}

Fig.~\ref{f:agepdf} shows the APDs of the Sun and HIP\,56948. For 
HIP 56948 we adopted errors from our differential analysis relative to the Sun
of 7\,K in $\tef$, 0.02\,dex in $\logg$,
and 0.01 dex in $\feh$, while for the Sun we adopted 
errors of 5\,K in $\tef$ and 0.005\,dex in $\logg$ and $\feh$. Although the errors 
for the Sun are overestimated, they allow us to obtain a smooth APD. 

Using our precisely determined 
stellar parameters for HIP\,56948, we derive a most probable age of 3.45\,Gyr.
The $1\,\sigma$ range of ages is 2.26--4.12 Gyr whereas the $2\,\sigma$ range of ages 
is 1.25--4.92\,Gyr. 
Notice that although the adopted He abundance may have some influence 
on the derived age, our small error bars in the stellar parameters of HIP 56948, 
rule out radically different He abundances. For relatively small changes in He,
the effect on the derived stellar age is relatively minor.
This is discussed in detail in appendix D, where stellar tracks with different He abundances are presented.
The mass is much better constrained; for HIP\,56948 we derive $1.020\pm0.016\,M_\odot$ ($2\,\sigma$ error).

In Fig.~\ref{f:agepdf} we also show the APD of HIP\,56948 assuming that its parameters 
were obtained using standard methods, for example obtaining $\tef$ from photometric data 
and $\logg$ from the \textit{Hipparcos} parallax. The standard method imply
an error of at least 100 K in $\tef$ due to uncertainties in the zero-point
of color-\teff relations \citep{cas10,mel10b}, while the error in 
Hipparcos parallax and typical errors in mass, \teff and V magnitude
imply a total error of about 0.07 dex in log $g$. For the error in [Fe/H] 
of a typical abundance analysis we adopt 0.05 dex. 
Clearly, in this case the age of HIP\,56948 is not well constrained. At most, 
we can say that the star is likely younger than about 8\,Gyr. Thus, 
another advantage of studying solar twins using very high quality spectroscopic data 
and strict differential analysis is that useful estimates of their ages can be 
obtained with the classical isochrone method, which can be a valuable asset for 
a variety of studies (e.g., Baumann et al. 2010).

The luminosity of HIP 56948 can be estimated from:
\begin{equation}
\log L/L_\odot = \log (M/M_\odot) \; - \; (\log g - 4.44) \; + \; 4 \, \log (\tef/5777)
\end{equation}

\noindent where $M$ is the stellar mass and $L$ the luminosity. 
Using our precise stellar parameters and mass we find
$L = 0.986 \pm0.051 \, L_\odot$. Thus, the luminosity of
HIP 56948 is essentially solar and therefore the extent of its 
habitable zone should be similar to the 
habitable region around the Sun \citep{kas93}.
From $L = 4\pi R^2 \sigma \tef^4$, we find 
a radius of $R = 0.987\pm0.023 \, R_\odot$, i.e., solar within the error bars.

\subsection{Further constraints on the age of HIP 56948}

Additional insight on the age of HIP 56948 can be obtained from
its chromospheric activity \citep{sod10}, lithium abundance \citep{don09,bau10}
and gyrochronology \citep{bar07}.

Determination of stellar ages based on chromospheric activity may only be
valid for solar type stars younger than $\sim$2 Gyr
\citep{pac04,pac09,zha11}, as older stars show a low
activity level which changes little with increasing age. 
Chromospheric activity is thus mainly
useful to distinguish young and old stars.
A measurement of chromospheric activity for HIP 56948 was obtained
by \cite{mel07} based on observations taken in April 2007,
resulting in a low chromospheric $S$-value of $S$ = 0.165 ($\pm$0.013)
in the Mount Wilson scale. Using another observation taken in
November 2007 we find a similar value, $S$ = 0.170  ($\pm$0.013).
Both values are as low as the the mean chromospheric activity 
index in the Sun, $<S_\odot>$ = 0.179 \citep{bal95}.
The low $S$-value of HIP 56948 suggests that it should be older than $\sim$2 Gyr.

\begin{figure}
\includegraphics[width=9.5cm]{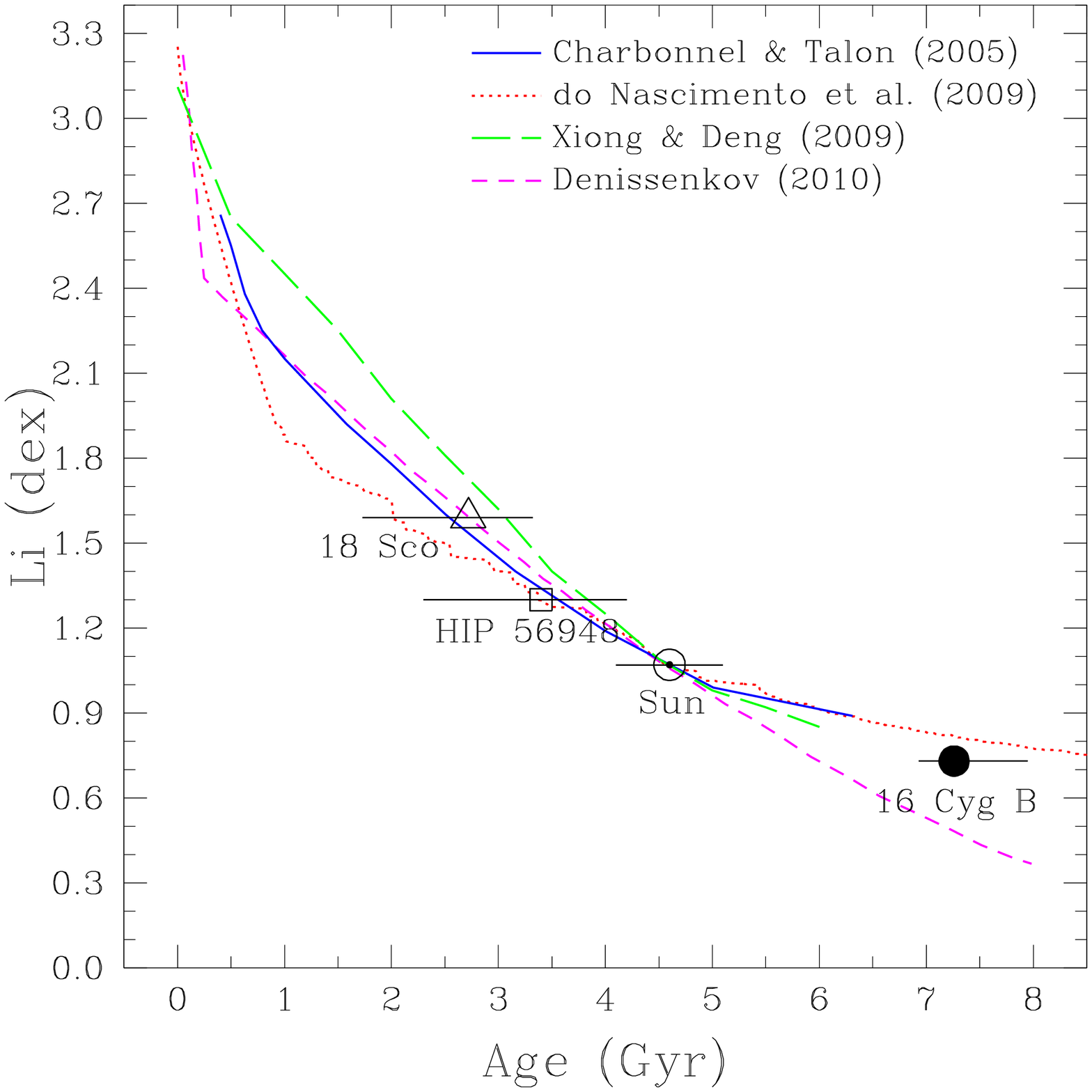}
\caption{Li vs. age for the Sun ($\odot$) and HIP 56948 (square) 
based on our NLTE Li abundances and isochrone ages,
and for 16 Cyg B (filled circle) and 18 Sco (triangle, 2.7 Gyr$^{-1.0}_{+0.6}$),
based on similar quantities by \cite{ram11} and Mel\'endez et al. (2012, in preparation).
The total error bar ($\pm \sigma$)
of the Li abundance is about the size of the symbols, while
the error bars in age are shown by horizontal lines.
For comparison we show the models by \cite{ct05,don09,xd09,den10}, shifted 
in Li abundance by 0.00, $-$0.03, $-$0.15, $-$0.05 dex, respectively, 
to reproduce our observed NLTE solar Li abundance.
The age of HIP 56948 based on Li tracks is in perfect 
agreement with the age obtained from isochrones.}
\label{f:ageli}
\end{figure}

Although it is well-known that lithium steeply decays with
age in very young stars \citep{sod10}, only recently it has been 
observationally shown that the decay continues for older ages \citep{mel10a,bau10},
as already predicted by several models of non-standard Li
depletion \citep{mon00,ct05,don09,xd09,den10}. Unfortunately we cannot
determine stellar ages with the required precision to discern
to what extent there is a spread (or not) in the age-lithium relation,
although the most precise values available \citep[this work;][Mel\'endez et al. 2012, in preparation]{ram11}
show relatively little dispersion (Fig.~\ref{f:ageli}).
Theoretical Li-age relations can be used to check
if a lithium age is compatible with the age determined from
isochrones. In Fig.~\ref{f:ageli} we compare several Li theoretical
tracks \citep{ct05,don09,xd09,den10} with our NLTE Li abundances and isochrone ages.
As can be seen, the agreement is excellent. If we were to derive an
age for HIP 56948 based on the theoretical Li-age relations, 
it would be 3.62$\pm$0.19 Gyr, i.e., almost the same age determined 
using isochrones. This comparison give us further
confidence that HIP 56948 is about 1 Gyr younger than the Sun.
Although the scatter of the Li age obtained using different
Li tracks is relatively small (0.19 Gyr), we conservatively assign 
an error of 1.0 Gyr to the Li-age, to take into account any possible
observational spread of the Li-age relation around solar age (Fig.~\ref{f:ageli}).

Stellar rotation can be used to estimate a 
rotational age \citep{bar07,sod10}.
Unfortunately there is no information on the rotation period
of HIP 56948 yet. Nevertheless, $v$ sin $i$ can give us an upper limit
on the rotation period, thus allowing to infer an upper limit on the age.
The determination of $v$ sin $i$ is based on the differential line broadening
between HIP 56948 and the Sun, as described in the appendix E. 

We infer for HIP 56948 $v$ sin $i$/$v$sin $i_\odot$ = 1.006 $\pm$ 0.014, or
$\Delta$ $v$ sin $i$ = +0.013 $\pm$ 0.026 km s$^{-1}$
(or $\pm$0.032 km s$^{-1}$ including the error in macroturbulence), i.e., 
HIP 56948 seems to have about the same rotation velocity as the Sun.
Within the uncertainties we infer that HIP 56948 cannot be
older than the Sun. Using the relation between rotation period
and age given in \cite{mel06} and \cite{bar07}, we find 
an upper limit of age  $\leq$ 4.7 Gyr. 

\begin{figure}
\includegraphics[width=9.5cm]{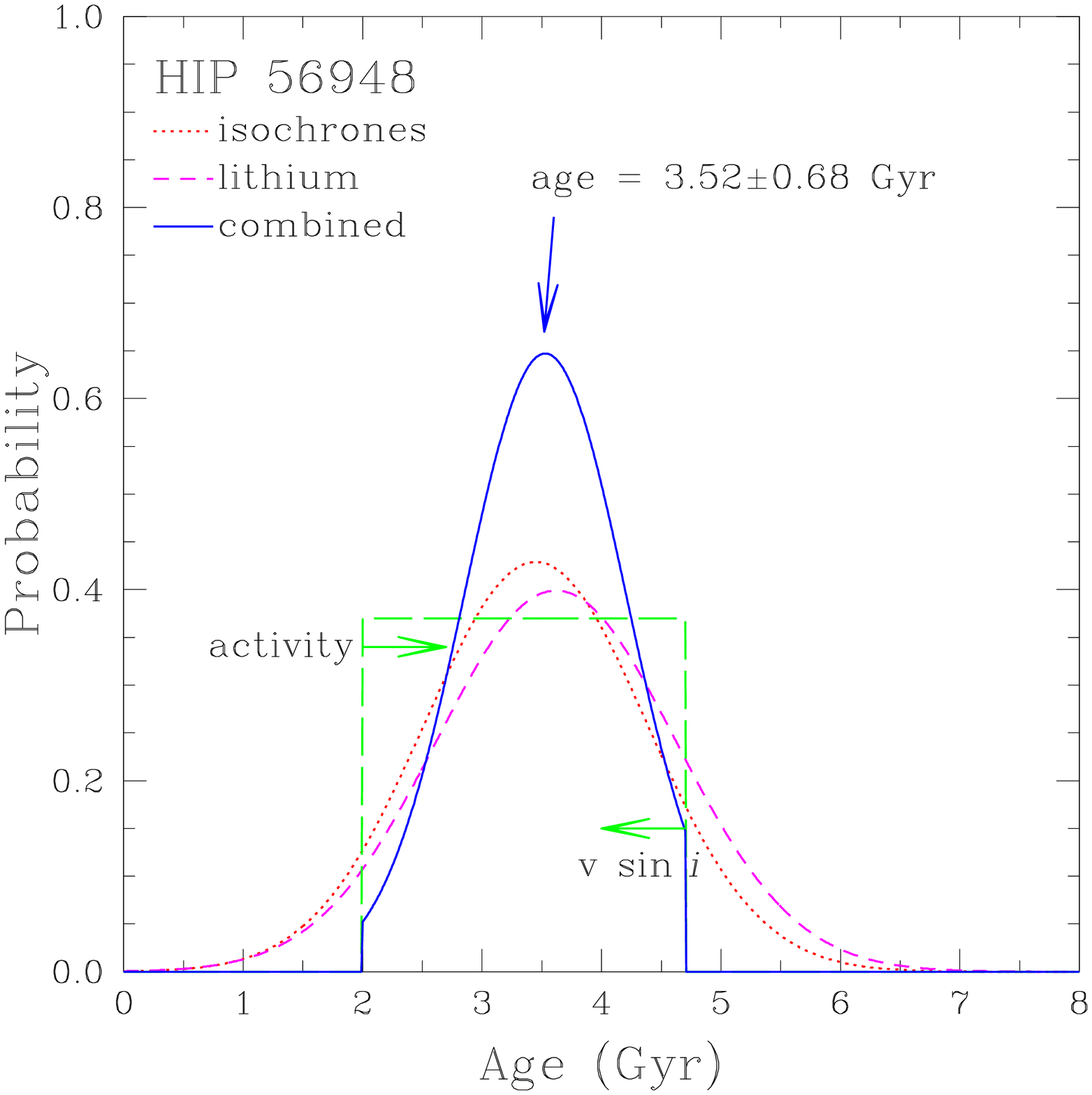}
\caption{Age probability distributions for HIP\,56948 based on
isochrones (dotted line) and lithium abundance (dashed line).
The limits on age imposed by chromospheric activity and by rotation (v sin$i$)
are also shown (long dashed lines). The combined age probability
distribution (solid line) is centered at 3.52 Gyr and has $\sigma$=0.68 Gyr.}
\label{age}
\end{figure}

Fig.~\ref{age} summarizes our findings on the age of
HIP 56948. Our precise stellar parameters and
differential isochrone analysis result in an age of $3.45\pm0.93$\,Gyr
(1-$\sigma$ error). The somewhat higher Li abundance of HIP 56948 
with respect to the Sun indicates an age of 
$\sim$3.62$\pm$1.00 Gyr. 
Chromospheric activity gives a lower limit of 2 Gyr,
while $v$ sin $i$ suggests an upper limit of 4.7 Gyr.
In Fig.~\ref{age} we show the combined age probability distribution.
Based on all the above indicators we suggest an age 
of 3.52$\pm$0.68 Gyr for HIP 56948.

Age is a key parameter for SETI programs \citep{tur03},
as stars only 1-2 Gyr old may not have had enough time
to develop complex life.
Life on Earth apparently appeared within the first billion year of 
the Earth's formation \citep{sch93,moj96,mck07,abr09}, but there is no
consensus on the exact date. Undisputed evidence for life 
can be traced back to about 2.7 Gyrs ago \citep[e.g., see review by][]{lop06}.
Yet, complex life only appeared about 0.5-1 Gyr ago
\citep{wra96,sei98,ras02,mar06}.
HIP 56948 is about one billion year younger than the Sun,
so assuming a similar evolution path as that of life on Earth,
complex life may be just developing (or already sprung if complexity
elsewhere can arise earlier than on Earth) in any hypothetical 
Earth that this star may host

\section{Conclusions}

We have shown that using spectra of superb quality coupled to a fully differential 
analysis of solar twins and the Sun, it is possible to achieve measurements errors as low as 0.003 dex 
for several elements (Si, Ca, Ti, Cr, Fe, Co, Ni).
Considering also the uncertainties in stellar parameters, we achieve an 
unprecedented accuracy of only $\sim$0.005 dex (1\,\%) 
in relative abundances for some elements and $<$ 0.01 dex for most elements.
This is almost one order of magnitude better than state-of-the-art works
in terms of absolute abundances \citep[e.g.,][]{asp05,asp09}.

The star HIP 56948 is remarkably similar to the Sun
in many different aspects. The  
effective temperature, log $g$, metallicity and microturbulence 
are very similar. The similarities also extend to its detailed
chemical abundance pattern. The volatile elements are in excellent
agreement, but the refractory elements are slightly (0.01 dex) 
more enhanced relative to the volatile elements in HIP 56948. 
From the comparison with the abundances of Earth-like and chondrite-like
material, we infer that about twice as much rocky material may have formed
around the Sun than around HIP 56948, albeit the amount of
Earth-like material is comparable for both stars ($\sim$2 M$_\oplus$).
The mass, luminosity and radius of HIP 56948 are essentially
solar within the uncertainties.
Lithium is severely depleted in HIP 56948, but not as much as in the Sun,
as expected for a solar twin somewhat younger than the Sun. 
Finally, our precise radial velocity data shows that 
the inner region around HIP 56948 is free from giant planets,
making thus more likely the existence of terrestrial planets
around this remarkable solar twin. Considering its similarities
to our Sun and its mature age, we urge the community to closely monitor 
HIP 56948 for planet and SETI searches, and to use other
techniques that could further our knowledge about HIP 56948,
such as for example asteroseismology, that have provided important
constraints for the solar twin 18 Sco \citep{baz11}.

The abundances anomalies we have discussed here cannot be explained by contamination 
from AGB stars, SNIa, SNII or HN (Sect. 4), or by Galactic chemical evolution processes 
or age effects \citep{mel09}. \cite{kis11} have shown that the peculiar abundance pattern 
cannot be attributed to line-of-sight inclination effects. Also, the abundance trend 
do not arise due to the particular reflection properties of asteroids (appendix B). 
Although the abundance peculiarities may indicate that the Sun was born in a massive 
open cluster like M67 \citep{one11}, this explanation is based on the analysis of only 
one solar twin. The Uppsala group is leading a high precision abundance study of other 
solar twins in M67, in order to confirm or reject this hypothesis. 
So far the best explanation for the abundance trend seems to be the formation of 
terrestrial planets. The Kepler mission should detect the first Earth-sized planets 
in the habitable zones of solar type stars. We look forward to 
use 8-10m telescopes to perform careful differential abundance analyses of those stars, 
in order to verify if our chemical signatures indeed imply rocky planets. 

\begin{acknowledgements}
The entire Keck/HIRES user community owes a huge debt to Jerry Nelson, Gerry Smith, Steve Vogt, 
and many other people who have worked to make the Keck Telescope and HIRES a reality 
and to operate and maintain the Keck Observatory. We are grateful to the
W. M.  Keck Foundation for the vision to fund the construction of the W.~M.~Keck Observatory.  
The authors wish to extend special thanks to those of Hawaiian ancestry on whose sacred 
mountain we are privileged to be guests.  Without their generous hospitality,
none of the observations presented herein would have been possible.
We thank Luca Casagrande for providing his estimates of \teff and log $g$
for HIP 56948, and Candace Gray and Caroline Caldwell for obtaining some
of the Tull spectra for precise radial velocities.
J.M. would like to acknowledge support from USP ({\em Novos Docentes}),
FAPESP ({\em 2010/17510-3}) and CNPq ({\em Bolsa de Produtividade}).
J.G.C. thanks NSF grant AST-0908139  for partial support. 
This work was performed in part (I.R.) under contract with the California Institute of Technology 
(Caltech) funded by NASA through the Sagan Fellowship Program.
ME, WDC and PJM were supported by NASA
Origins of Solar Systems grant NNX09AB30G.
This publication has made use of the SIMBAD database, operated at CDS,
Strasbourg, France.
\end{acknowledgements}

\Online

\begin{appendix} 
\section{Is the T$_{\rm cond}$-abundance trend real ?}

\begin{figure}
\resizebox{\hsize}{!}{\includegraphics{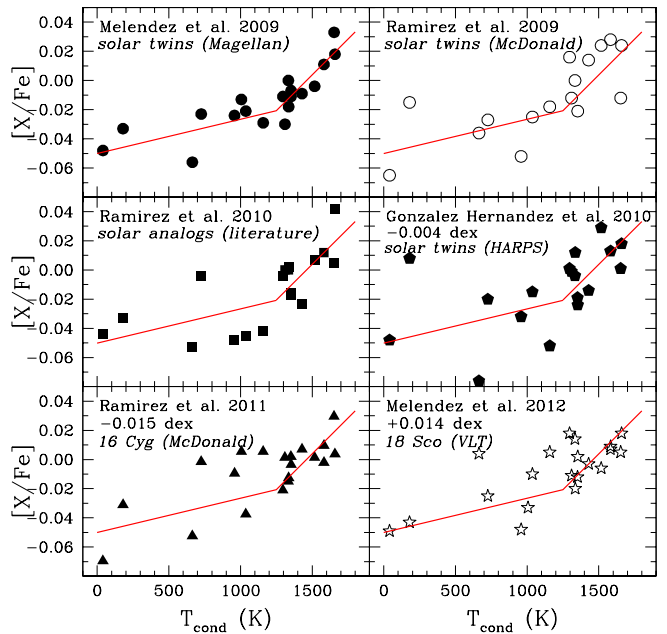}}
\caption{
[X/Fe] ratios (from carbon to zinc) vs. condensation temperature for different samples of solar
twins and solar analogs. The solid line represents the mean trend found by \cite{mel09}.
{\it upper-left} (filled circles): Southern sample of solar twins by \cite{mel09}; 
{\it upper-right} (open circles): Northern solar twin sample by \cite{ram09};
{\it middle-left} (squares): average of six different literature samples of solar analogs \citep{ram10};
{\it middle-right} (pentagons): average of 15 solar twins in the HARPS sample of \cite{gh10},
after a shift of $-$0.004 dex;
{\it lower-left} (triangles): average of the pair of solar analogs 16 Cyg A and B \citep{ram11},
after a shift of $-$0.015 dex;
{\it lower-right} (stars): abundance pattern of the solar twin 18 Sco (Mel\'endez et al. 2012, in preparation),
after a shift of $+$0.014 dex;.
}
\label{tclit}
\end{figure}

We are just starting the era of high precision (0.01 dex) abundances studies, therefore the casual 
reader may question how real or universal is the abundance trend ($<$solar twins$>$ - Sun) with condensation temperature. 
The trend was first found by \cite{mel09}, who determined a Spearman correlation coefficient of $r_S$ = $+$0. 91 
and a negligible probability of only $\sim$ 10$^{-9}$ of this trend to happen by pure chance. These results, 
based on Southern solar twins observed at the Magellan telescope, are reproduced in the left-upper panel 
of Fig. \ref{tclit}, where the average abundance ratios of the solar twins is plotted against condensation temperature. 
Additional independent works are also shown in Figure \ref{tclit}, where a line representing the mean 
trend found by \cite{mel09}, is superimposed upon the different samples. 

The independent study by \cite{ram09}, using McDonald data of Northern solar twins, 
follows the same trend (Fig. \ref{tclit}, upper-right panel), as well as the average 
of six independent samples \citep[][Bensby et al., in preparation]{red03,all04,takeda07,nev09,gon10} of solar analogs in the literature \citep{ram10}, 
as shown in the left-middle panel of Fig. \ref{tclit}. 

The revision and extension \citep{gh10} of the abundance analysis by \cite{nev09} of 
the HARPS high precision planet survey, also follows the same trend, 
except for a minor global shift of only $-$0.004 dex, as illustrated in the right-middle panel of Fig. \ref{tclit}.
In this panel we show the average abundance ratios\footnote{The individual 
abundance ratios with errors larger than 0.1 dex were discarded from the data of \cite{gh10}
when computing the average [X/Fe] values.}
of 15 HARPS solar twins with \tsin, log $g$ and [Fe/H] within $\pm$100 K, $\pm$0.1 dex and $\pm$0.1 dex 
of the Sun's stellar parameters. 
The agreement between the solar twin pattern of \cite{gh10} and the mean trend of
\cite{mel09} is good, except for the element O and to a lesser extent for S, 
for which it is difficult to determine precise abundances. In particular,  
notice that \cite{gh10} derived oxygen abundances from the [OI] 630nm line, 
which is badly blended with NiI. Also, notice that the [OI] feature is the 
weakest employed by them. 
Since their oxygen abundance is based on a single line, which is the weakest of all 
features analyzed by them, and that [OI] is blended with NiI, it is natural to 
expect that the O abundances in \cite{gh10} have the largest uncertainties.
Furthermore, the HARPS spectra taken for planet hunting shows some contamination from the 
calibration arc, so that abundances derived from only one single feature should be taken with care.

Regarding the analysis of individual stars, most of them also display the abundance trend,
as shown for example for four solar twins in Fig. 2 of \cite{ram09} and 11 solar twins in Fig. 4 of \cite{mel09}.
Two new examples are shown in Fig. \ref{tclit}. In the bottom-left panel we shown the 
average abundance of the pair of solar analogs 16 Cyg A and B \citep{ram11}, 
based on high resolution (R = 60,000) and high S/N ($\sim$400) McDonald observations. 
As can be seen, this pair also follows the abundance 
trend, after a minor shift of $-$0.015 dex. In the bottom-right panel we show the abundance 
ratios of the solar twin 18 Sco (Mel\'endez et al. 2012, in preparation), based on high quality
(R = 110,000, S/N $\sim$ 800) UVES/VLT data. 
It is clear that the abundance trend is also followed by 18 Sco, 
after a shift of only +0.014 dex in the abundance ratios.
Similar results are obtained using HIRES/Keck data (Mel\'endez et al. 2012, in preparation;
see also appendix B).

Thus, all recent high precision abundance studies based on different samples of 
solar twins and solar analogs in the Southern and Northern skies, using different instrumentation 
(Tull Coude Spectrograph at McDonald, MIKE at Magellan, HARPS at La Silla, UVES at the VLT, HIRES at Keck), 
all show the abundance trend. In conclusion, it seems that the reality of the 
abundance trend found by \cite{mel09} and \cite{ram09}, is well established.

\end{appendix}

\begin{appendix} 
\section{Test of our precision using the asteroids Juno and Ceres}

The referee suggested that we test our method using observations of two asteroids of different properties, 
obtained with the same instrument and setup, in order to show whether our very small 
standard errors ($\sim$0.005 dex) are adequate to estimate the observational uncertainties, as well 
as to look for potential systematic problems with the asteroid Ceres. Although we have not acquired such data yet, 
we do have observations of the solar twin 18 Sco and two different asteroids observed with 
different instruments: high quality UVES spectra of 18 Sco and the asteroid Juno 
(R = 110,000 and S/N $\sim$ 800) and high quality HIRES spectra of 18 Sco and 
the asteroid Ceres (R = 100,000, S/N $\sim$ 400). Juno is a S-type asteroid 
and Ceres is a C-type asteroid \citep[e.g., ][]{dem09}, therefore they have very different spectral 
properties and the relative analysis of 18 Sco to both Juno and Ceres should reveal 
if there is any problem in using their reflected solar light.

\begin{figure}
\resizebox{\hsize}{!}{\includegraphics{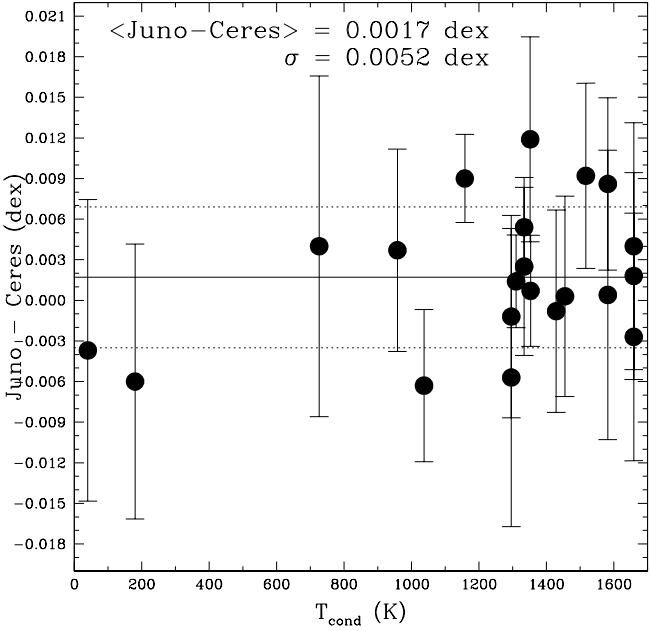}}
\caption{
Differences between the solar abundances obtained with the asteroids 
Juno and Ceres, obtained through 
(Juno - Ceres) = [X/H](18 Sco - Ceres) - [X/H](18 Sco - Juno).
The solid line shows the mean difference and the dotted lines
show the element-to-element scatter.
}
\label{hiresuves}
\end{figure}

The analysis has been performed as described in Sect. 3. 
Our preliminary LTE results for the stellar parameters of 
18 Sco are \teff = 5831$\pm$10 K, log $g$ = 4.46$\pm$0.02 dex, [Fe/H] = 0.06$\pm$0.01 dex.
Full details of the abundance analysis will be published elsewhere. 
In Fig.\ref{hiresuves} we show the difference between the [X/H] ratios 
obtained in 18 Sco using the Ceres and Juno asteroids, 
[X/H]$_{\rm 18 Sco - Ceres}$ - [X/H]$_{\rm 18 Sco - Juno}$, or in other words the 
abundance difference (Juno - Ceres). Notice that the same set of lines was 
used for both analyses. The error bars shown in Fig.\ref{hiresuves}  are the 
combined error bar based on the standard error (s.e.) of each analysis, i.e, 
error = $\sqrt{s.e.^2_{\rm 18 Sco - Ceres} + s.e.^2_{\rm 18 Sco - Juno} }$.

As can be seen, the standard errors fully explain the small deviations of 
the (Juno - Ceres) abundance ratios. 
The mean difference $<$Juno - Ceres$>$ is only 0.0017 dex, 
and the element-to-element scatter is only 0.0052 dex, meaning
that each of the individual analyses should have typical errors of about 0.003 - 0.004 dex.
The agreement is very satisfactory considering the different instrumentation employed and that the 
comparison between Juno and Ceres is done through a third object 
(the solar twin 18 Sco). Also, notice that there is no meaningful trend with 
condensation temperature. The test performed here strongly supports 
for our high precision and removes the possibility that the abundance trend 
may arise due to the particular properties of asteroids.

\end{appendix}

Besides the potential applications of high precision differential abundance 
techniques to study the star-planet connection, these
techniques are also giving new insights in other areas. 
\cite{ns10} achieved uncertainties of 0.03 dex in [Mg/Fe] and
only 0.02 dex in both [Ca/Fe] and [Ti/Fe], 
showing a clear separation of the halo into two distinct populations with different [$\alpha$/Fe] ratios. 
Regarding globular clusters, \cite{mc09} have shown that CN-weak giants in M71
show a star-to-star scatter in [O/Fe] and [Ni/Fe] of only 0.018 dex, while [Mg/Fe] and [La/Fe] 
have a scatter of 0.015 dex. The star-to-star scatter in metallicity ([Fe/H]) is only 0.025 dex. 
Even a lower star-to-star scatter is found among ``globular cluster star twins'' (stars within $\pm$100 K of a globular 
cluster standard star) of NGC 6752. Using superb spectra (R = 110,000; S/N = 500) obtained
with UVES on the VLT \citep{yon03,yon05} and applying similar techniques to those presented in Sect. 3, Yong et al. (2012, in preparation) 
have found an unprecedentedly low star-to-star scatter of only 0.003 dex in the 
iron abundances among NGC 6752 star twins, revealing chemical homogeneity in this cluster
at the 0.7\% level.

\begin{appendix} 
\section{Determination of stellar parameters}

\begin{figure}
\resizebox{\hsize}{!}{\includegraphics{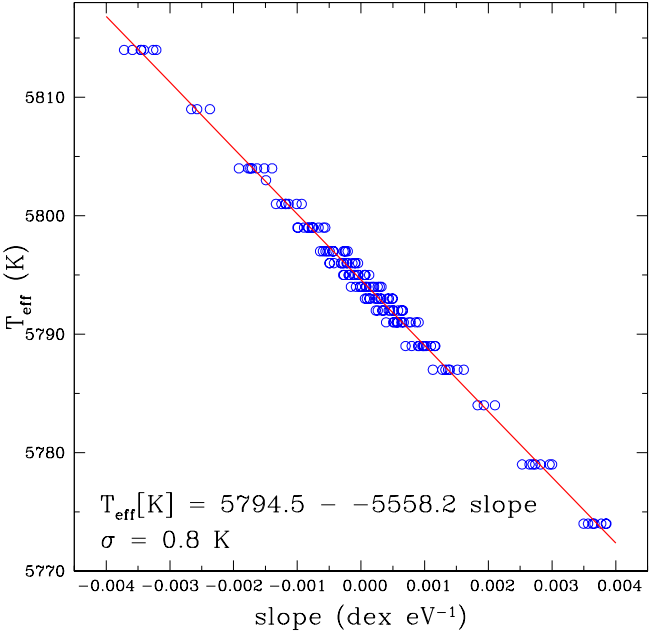}}
\caption{
{\teff} as a function of the slope = $d(\delta A_i^{\rm FeI}) / d(\chi_{\rm exc})$
(see eq. 2), for a range of surface gravities (4.40 dex $\leq$ log $g$ $\leq$ 4.52 dex),
corresponding to a scatter of only 0.8 K in {\tsin}.
The result of different model atmospheres are shown as open circles.
The line represents a linear fit.
}
\label{teff}
\end{figure}

\begin{figure}
\resizebox{\hsize}{!}{\includegraphics{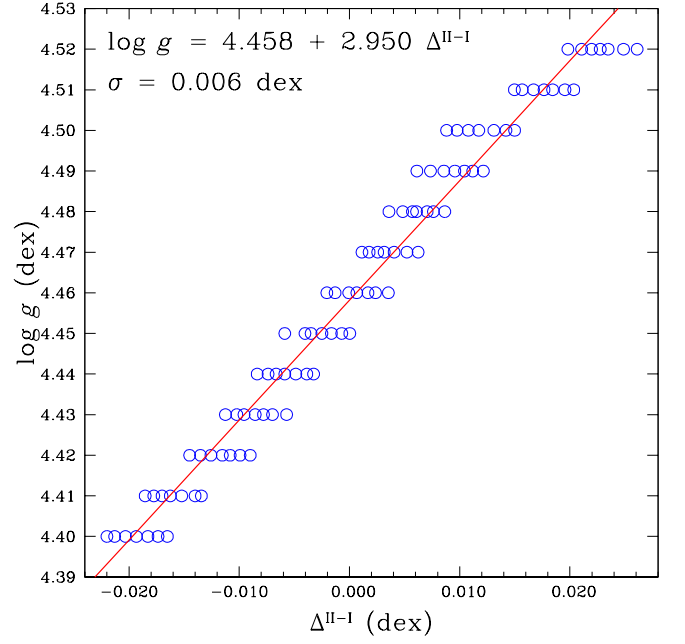}}
\caption{
Surface gravity vs. $\Delta^{\rm II - I} = 
( 3\Delta^{\rm FeII-FeI} \; + \; 2 \Delta^{\rm TiII-TiI} \; + \; \Delta^{\rm CrII-CrI}) / 6$ (see eq. 3).
The spread of the circles correspond to a range of effective temperatures (5791 K $\leq$ {\teff} $\leq$ 5797 K),
implying in a scatter of 0.006 dex in log $g$. The line represents a linear fit.
}
\label{logg}
\end{figure}

As mentioned in Sect. 3, the excitation and ionization equilibrium do not depend only on \teff and log $g$, 
respectively. There is some dependence with other stellar parameters, but to a much lesser extent, 
such that a ``unique'' solution can easily be obtained after a few iterations. 
In practice, considering the weak degeneracies, a first guess of the effective temperature can be 
obtained computing the slope at three different \teff (e.g., in steps of 50 K and at fixed solar log $g$) 
at the best microturbulence velocity (at a given \teff and log $g$). Then a linear fit is 
performed to \teff vs. slope (see Fig. \ref{teff}) to find the effective temperature at slope = 0. 
Then, for this \teff we can run three models with different log $g$ (e.g., in steps of 0.05 dex in log $g$) 
in order to find the best surface gravity, by fitting  log $g$ vs. $\Delta^{\rm II-I}$ (see Fig. \ref{logg}), 
and for each model the microturbulence is obtained. This leads to the first guess of 
\tsin, log $g$ and $v_t$. Further iterations at smaller steps (down to 1 K in \teff, 0.01 dex in log $g$
and 0.01 km s$^{-1}$ in $v_t$) 
can quickly lead to the best solution that simultaneously satisfy the conditions of differential 
spectroscopic equilibrium (eqs. 2-4).

In Fig. \ref{teff} we show that the excitation equilibrium provides a precise \tsin.
In this figure, effective temperature is plotted versus the slope = $d(\delta A_i^{\rm FeI}) / d(\chi_{\rm exc})$. 
As can be seen, there is a clear linear relation between \teff and the slope, 
with some minor spread of only 0.8 K due to a range in adopted surface gravities.

Regarding the ionization equilibrium, we show in Fig. \ref{logg} the dependence between 
surface gravity and $\Delta^{\rm II - I}$ (see eq. 3). A linear fit represents well this relation, 
with a scatter in log $g$ of only 0.006 dex for models in a range of effective temperatures.

\begin{figure}
\resizebox{\hsize}{!}{\includegraphics{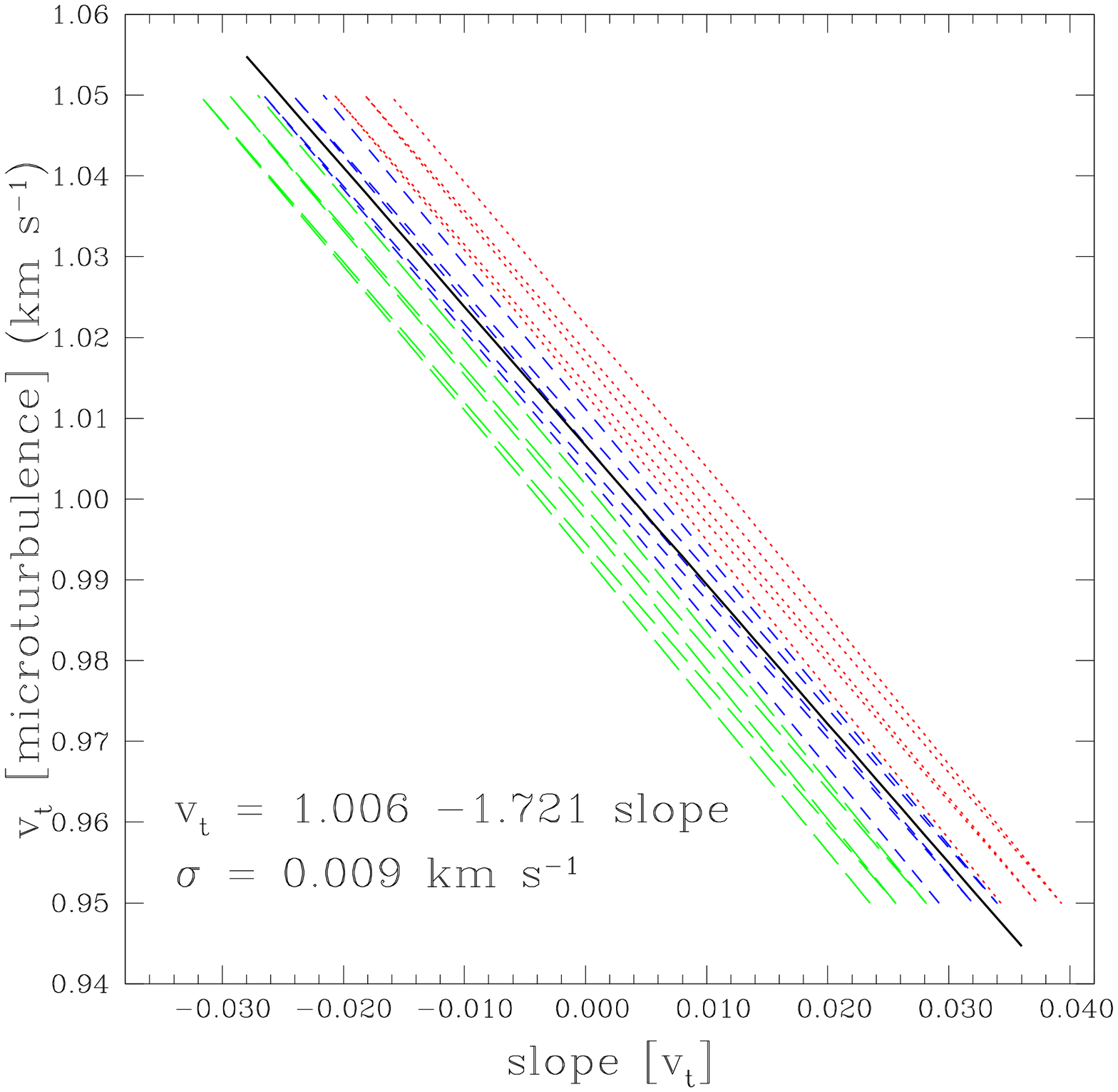}}
\caption{
Microturbulence velocity vs. slope $d(\delta A_i^{\rm FeI}) / d(EW_r)$ (see eq. 4).
The dotted, dashed and long dashed lines are for \teff = 5791, 5794, 5797 K, respectively,
and the spread shown for each line style is due to a range in log $g$ (4.40 dex $\leq$ log $g$ $\leq$ 4.52 dex).
This spread in \teff and log $g$ corresponds to a scatter of 0.009  km s$^{-1}$ in $v_t$.
The solid line represents a linear fit.
}
\label{vmicro}
\end{figure}

In Fig. \ref{vmicro} we show the linear dependence between microturbulence velocity 
and the slope $d(\delta A_i^{\rm FeI}) / d(EW_r)$. For a given model, 
$v_t$ could be constrained to within 0.0004 km s$^{-1}$. 
A range in \teff (5791 K $\leq$ {\teff} $\leq$ 5797 K) and log $g$ (4.40 dex $\leq$ log $g$ $\leq$ 4.52 dex) 
imply in a scatter of only 0.009 km s$^{-1}$ in $v_t$.

Given the above dependences, the stellar parameters 
\tsin, log $g$ and $v_t$ must be iteratively modified until the 
spectroscopic equilibrium conditions (equations 2-4) are satisfied
simultaneously. 
Since the degeneracy is relatively small, the final solution 
(\tsin/log $g$/$v_t$ = 5794 K/4.46 dex/1.00 km s$^{-1}$) is very close 
to the independent solutions shown in 
Figs. \ref{teff} - \ref{vmicro} (5794.5 K/4.458 dex/1.006 km s$^{-1}$). 

In order to check how unique is the derived final solution,
we have run over 200 models with different stellar parameters,
with a very fine grid (steps of only 1 K in \teff and 0.01 dex in log $g$) 
near our best solution. We then evaluated how close to zero
are the slope in \teff (eq. 2) and the ionization equilibrium parameter $\Delta^{\rm II-II}$
(eq. 3). The following quantity is evaluated for each model,

\begin{equation}
TG = ( |slope/error| + |\Delta^{\rm II-II}/error| )/2.
\end{equation}

\begin{figure}
\resizebox{\hsize}{!}{\includegraphics{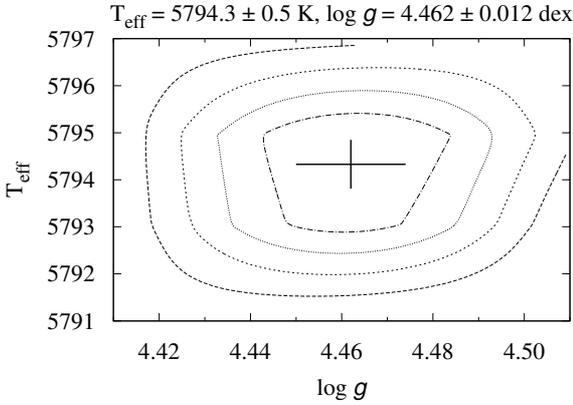}}
\caption{
Contour plot of the parameter $TG$ (eq. C.1), which evaluates how good is 
the differential spectroscopic equilibrium. The minimum is shown by a cross at
\teff =  5794.3$\pm$0.5 K and log $g$ = 4.462$\pm$0.012 dex, 
which is in excellent agreement with our adopted solution.
The contour levels increase in steps of $\Delta TG$ = 0.1 from the minimum.
}
\label{contour}
\end{figure}

The model showing the lowest $TG$ value would be the best
spectroscopic solution, which in our case is obtained for \teff = 5794 K,
log $g$ = 4.46 dex, and $v_t$ = 1.00 km s$^{-1}$. 
A contour plot for the $TG$ parameter is shown in Fig.\ref{contour}. 
Besides the best solution at \teff  = 5794 K and log $g$ = 4.46 dex, 
there are a few other nearby plausible solutions, with a mean 
value at  \teff =  5794.3$\pm$0.5 K and log $g$ = 4.462$\pm$0.012 dex,
shown by a cross in Fig.\ref{contour}. 
Our grid samples a much larger coverage than that shown in Fig. \ref{contour}, 
and we have verified that the best solution indeed represents a global minimum,
i.e, there is no other solution that can simultaneously satisfy the 
conditions of differential spectroscopic equilibrium. Thus, 
within the error bars our solution is ``unique''.

\end{appendix}

\begin{appendix} 
\section{Helium abundance and the age and log $g$ of HIP 56948}
Since our stellar parameters are very precise, actually the He abundance in 
HIP 56948 cannot be arbitrarily different from the solar He abundance. 
For example, an evolutionary track computed with a He abundance 5\% higher than solar, would shift 
the \teff by about +74 K at the same log $g$, i.e., a change 10 times 
larger than our error bar in \tsin, leading thus to no plausible solutions. 
We are currently building an extensive grid of models with He as a free parameter, 
using the Dartmouth stellar evolution code \citep{cha01,gue92}, which is based upon 
the Yale stellar evolution code. 

\begin{figure}
\resizebox{\hsize}{!}{\includegraphics{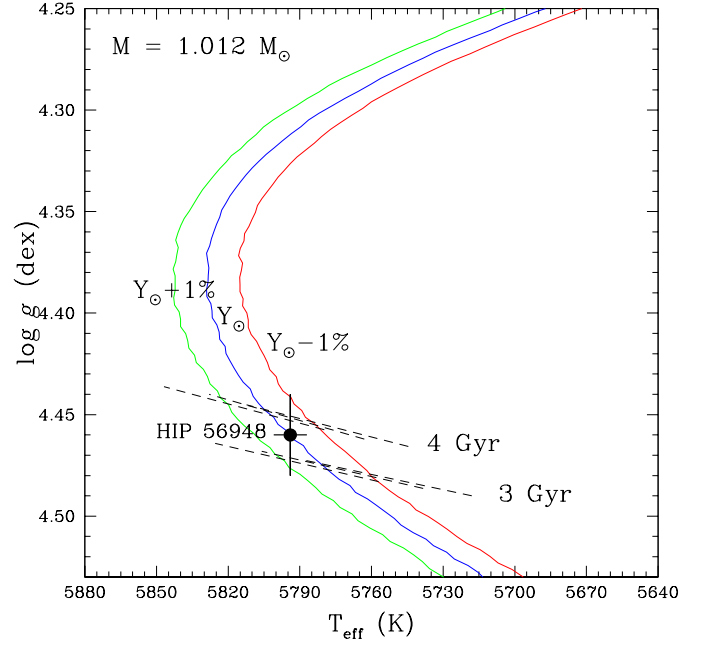}}
\caption{
Solar metallicity evolutionary tracks for 1.012 M$_\odot$ (solid lines)
at three different He abundances: solar and $\pm$1\% solar. 
Isochrones at 3 and 4 Gyr are plotted with dashed lines.
The position of HIP 56948 and error bars in \teff and log $g$ are also shown.
}
\label{helium}
\end{figure}

In Fig. \ref{helium}, we show evolutionary tracks for M = 1.012 M$_\odot$ (solid lines), which
is the best mass found for HIP 56948 using the Dartmouth tracks 
adopting the He solar abundance. These models were computed
at three different helium abundances, solar and $\pm$1\% solar. 
We also show isochrones at 3 and 4 Gyr for different He abundances.
The error bars in log $g$  and \teff put constraints on the
He abundance, which can not be radically different from solar. Notice that the isochrones run parallel 
to each other and with only a minor shift for a change of $\pm$1\% in the He abundance, 
resulting thus in about the same central solution for age, independent of the adopted He content.
Therefore thanks to our small error bars in stellar parameters we can
put stringent constraints on the age of HIP 56948.
Interestingly, there is a degeneracy between He and mass, 
although below our 2\% error bar in mass. 

Another example where the adoption of a somewhat different He abundance
did not affect much the stellar age is for the pair of solar analogs 16 Cyg A and B.
For this pair, asteroseismology have recently constrained the ages of these stars, which
are in excellent agreement with those derived from our isochrone technique,
despite the somewhat different adopted He abundances. 
Based on three months of almost uninterrupted Kepler observations, \cite{met12} obtained 
an age = 6.8$\pm$0.4 Gyr for their optimal models, in excellent agreement 
with an age = 7.1$\pm$0.4 Gyr derived from our isochrone technique for 
the 16 Cyg pair \citep{ram11}.

The effect of changing He by $\pm$1\% (in Y) in our model atmospheres has 
also a minor impact on the derived spectroscopic log $g$. 
As shown by \cite{str82}, for solar type dwarfs the change in log $g$ due to 
a change in the helium to hydrogen ratio ($y = N_{\rm He}/N_{\rm H}$) is:

\begin{equation}
{\rm log} \,  g = {\rm log} \, g\,' + {\rm log} \, \left[\frac{(1 + 4 y\,')(1+y)}{(1+4y)(1+y\,')}\right].
\end{equation}

\cite{lin11b} have shown that the relation above is also adequate for giant stars.
For a change of +1\% in Y in HIP 56948, the predicted change in log $g$ is only -0.001 dex, 
which is well below our error bar in log $g$ (0.02 dex). Thus, assuming that the 
He abundance of HIP 56948 is not radically different from solar, 
our derived spectroscopic log $g$ value is essentially unaffected.

\end{appendix}

\begin{appendix} 
\section{$v$ sin $i$ and macroturbulence velocity}

We have determined $v$ sin $i$  from the differential line broadening
(HIP 56948 - Sun). Naively we could be 
assuming an identical macroturbulence for both stars, but
at a given luminosity class, macroturbulence seems a smooth 
function of temperature \citep[e.g.,][]{saa97,gra05,val05}, so neglecting
this effect can lead to a slight overestimation of $v$ sin $i$ in HIP 56948 because it
is hotter than the Sun and therefore the contribution of $v_{\rm macro}$ to
the line broadening in HIP 56948 should be slightly larger.

The trend of macroturbulence velocity with \teff described
by \cite{gra05} for main sequence stars\footnote{For subgiants, giants and luminous giants,
our fits can be found in \cite{hm07}} can be fitted by

\begin{equation}
v_{\rm macro} = 13.499 - 0.00707 \, \tef + 9.2422 \times 10^{-7}  \tef^2.
\end{equation}

\noindent A similar correlation was advocated by \cite{val05} (after normalization
to $v_{\rm macro}^\odot$ = 3.50 km s$^{-1}$, which is the value obtained for the Sun
using Gray's relation):

\begin{equation}
v_{\rm macro} = 3.50 + (\tef - 5777)/650.
\end{equation}

\noindent Finally, the mean relation (active and non-active stars) obtained
by \cite{saa97}, after transforming (B-V) to \teff \citep{val05}
and normalizing it to $v_{\rm macro}^\odot$ = 3.50 km s$^{-1}$, is:

\begin{equation}
v_{\rm macro} = 3.50 + (\tef - 5777)/388.
\end{equation}

\noindent The first two relations are valid for $\sim$ 5000-6500 K, while
the last relation is valid for $\sim$ 5000-6100 K.
On average, the above relations predict a differential (HIP 56948 - Sun) 
$\Delta v_{\rm macro} = 0.044 \pm 0.018$ km s$^{-1}$.

In order to determine $v$ sin $i$ we selected 
19 lines in the 602-682 nm region, although essentially similar results are
obtained (albeit with even so slightly larger errors) when 50 lines
covering the 446-682 nm region are used. 
First, we performed spectral synthesis of selected lines, in order
to calibrate the relation between line width (in \AA) and total broadening (in km s$^{-1}$).
Then, we estimated the total broadening using a much larger set of lines,
and obtained $v$sin$i$ after subtracting 
both the instrumental and the macroturbulence broadening.

After taking into account the somewhat higher macroturbulence
velocity of HIP 56948, we find $v$ sin $i$/$v$sin $i_\odot$ = 1.006 $\pm$ 0.014, or
$\Delta$ $v$ sin $i$ = +0.013 $\pm$ 0.026 km s$^{-1}$
(or $\pm$0.032 km s$^{-1}$ including the error in macroturbulence), i.e., 
HIP 56948 seems to have about the same rotation velocity as the Sun, or 
rotating slightly faster, although it is unclear how much
faster due to the uncertain sin $i$ factor.

\end{appendix}

\clearpage

\scriptsize
\begin{table}
\caption{McDonald radial velocity measurements for HIP 56948}
\label{tabradvel}
\centering 
\renewcommand{\footnoterule}{}  
\begin{tabular}{ccccccllll} 
\hline
\hline 
{BJD}  & Velocity  & Error  \\
{}     & (m s$^{-1}$) & (m s$^{-1}$)  &  \\
\hline
      2454250.742247 &     1.86 &   3.51\\
      2454251.762237 &     6.18 &   6.37\\
      2454347.603043 &    -0.26 &   8.35\\
      2454822.022016 &    -3.17 &   4.48\\
      2455285.780573 &     6.53 &   6.00\\
      2455585.972936 &    -5.06 &   4.07\\
      2455643.770312 &    -0.17 &   4.67\\
      2455667.816687 &   -16.11 &   4.17\\
      2455990.845569 &    10.21 &   5.51\\
\hline                                 
\end{tabular}
\end{table}

\scriptsize
\begin{table}
\caption{Keck radial velocity measurements for HIP 56948}
\label{tabradvelkeck}
\centering 
\renewcommand{\footnoterule}{}  
\begin{tabular}{ccccccllll} 
\hline
\hline 
{BJD}  & Velocity  & Error  \\
{}     & (m s$^{-1}$) & (m s$^{-1}$)  &  \\
\hline
  2455610.068981   &  -2.77   & 3.24 \\
  2455611.013099   &  -6.06   & 3.73 \\
  2455611.100610   &  -9.49   & 3.38 \\
  2455611.148864   & -10.18   & 3.83 \\
  2455766.748023   &   5.54   & 2.20 \\
  2455767.742970   &   4.39   & 2.65 \\
  2455767.746153   &   0.82   & 3.78 \\
  2455767.749454   &   3.74   & 2.00 \\
  2455935.157734   &  -0.35   & 2.43 \\
  2455935.160951   &   1.21   & 1.89 \\
  2455935.163959   &   0.67   & 2.34 \\
  2455936.135676   &  -4.52   & 3.28 \\
  2455936.138532   &  -5.07   & 1.98 \\
  2455936.141412   &  -8.27   & 2.68 \\
  2455937.130793   &   0.19   & 2.85 \\
  2455937.133891   &  -1.58   & 2.46 \\
  2455937.137006   &  -1.07   & 2.54 \\
  2455938.111144   &   3.23   & 3.62 \\
  2455938.114027   &   3.34   & 2.50 \\
  2455938.116890   &  -2.13   & 4.05 \\
  2455962.089413   &   1.70   & 2.95 \\
  2455962.092257   &   0.20   & 2.62 \\
  2455962.095094   &   0.85   & 1.84 \\
  2455963.104952   &   3.75   & 3.12 \\
  2455963.107611   &   5.20   & 1.86 \\
  2455963.110271   &   1.18   & 2.34 \\
  2455964.037962   &   1.92   & 2.66 \\
  2455965.071634   &   3.50   & 1.98 \\
  2455965.075176   &   4.35   & 3.92 \\
  2455965.078440   &   5.71   & 2.31 \\
\hline                                 
\end{tabular}
\end{table}

\addtocounter{table}{1}
\scriptsize
\begin{table}
\caption{Stellar parameters and Li abundance of HIP 56948 relative to the Sun (HIP 56948 - Sun)}
\label{tabparam}
\centering 
\renewcommand{\footnoterule}{}  
\begin{tabular}{ccccccllll} 
\hline
\hline 
{$\Delta$\tsin}  & $\Delta$ log $g$ & $\Delta$[Fe/H]  & $\Delta$v$_t$ & $\Delta$Li (NLTE) & $\Delta v$ sin $i$  & method & reference \\
{(K)}     &   (dex) & (dex)  & (km s$^{-1}$) & (dex) & (km s$^{-1}$) &        &  \\
\hline
17$\pm$7  & +0.02$\pm$0.02 & +0.02$\pm$0.01 & +0.01$\pm$0.01 & 0.23$\pm$0.05 & +0.01$\pm$0.03 & spectroscopy & This work   \\
\hline
26$\pm$70 & -0.07$\pm$0.07 &                &                &                &        & IRFM, Hipparcos & L. Casagrande (priv. communication) \\
26$\pm$63 &               &                &                &                &        & IRFM          & \cite{cas11} \\
24$\pm$25\tablefootmark{a} &    &          &                 &                &        & IRFM         & \cite{cas10} \\
17$\pm$5 & +0.01$\pm$0.01 & +0.02$\pm$0.01 & -0.01$\pm$0.01  & 0.22           & +0.05  & spectroscopy & \cite{tak09} \\
3$\pm$5  & -0.02$\pm$0.01 & +0.01$\pm$0.01 & -0.01$\pm$0.04  &                &        & spectroscopy & \cite{tak09} \\
60$\pm$56 & +0.03$\pm$0.08 & +0.04$\pm$0.03 &                & 0.22$\pm$0.07  & 0.0$\pm$0.1 & spectroscopy & \cite{ram09} \\
5$\pm$36 & -0.04$\pm$0.05 & +0.01$\pm$0.02 & +0.01$\pm$0.06  & -0.02$\pm$0.13 & 0.0$\pm$0.1 & spectroscopy & \cite{mel07} \\
-2$\pm$52 &               &                &                 &                &        & photometry & \cite{mas06} \\
\hline                                 
\end{tabular}
\tablefoot{
\tablefoottext{a}{Error bar based only on photometric errors}
}
\end{table}

\clearpage

\scriptsize
\begin{table}
\caption{Atomic and molecular line list in the vicinity of the Li lines}
\label{listli}
\centering 
\renewcommand{\footnoterule}{}  
\begin{tabular}{llll} 
\hline      
\hline 
{Wavelength}  & Species  & $\chi_{\rm exc}$ & log $gf$ \\
{~~~~~~~\AA}         &          & (eV)                 & (dex)    \\
\hline
6706.5476 & CN     & 3.13  & -1.359 \\  
6706.5665 & CN     & 2.19  & -1.650 \\  
6706.657  & CN     & 0.860 & -2.993 \\  
6706.658  & CN     & 0.614 & -3.622 \\  
6706.728  & CN     & 0.625 & -2.400 \\  
6706.7329 & CN     & 0.870 & -1.768 \\  
6706.8440 & CN     & 1.96  & -2.775 \\  
6706.8626 & CN     & 2.07  & -1.882 \\  
6706.880  & Fe II  & 5.956 & -4.103 \\  
6707.00   & Si I   & 5.954 & -2.56  \\  
6707.172  & Fe I   & 5.538 & -2.810 \\  
6707.2052 & CN     & 1.97  & -1.222 \\  
6707.272  & CN     & 2.177 & -1.416 \\  
6707.2823 & CN     & 2.055 & -1.349 \\  
6707.300  & C$_2$  & 0.933 & -1.717 \\  
6707.3706 & CN     & 3.05  & -0.522 \\  
6707.433  & Fe I   & 4.608 & -2.25  \\  
6707.460  & CN     & 0.788 & -3.094 \\  
6707.461  & CN     & 0.542 & -3.730 \\  
6707.4695 & CN     & 1.88  & -1.581 \\  
6707.473  & Sm II  & 0.933 & -1.91  \\  
6707.548  & CN     & 0.946 & -1.588 \\  
6707.5947 & CN     & 1.89  & -1.451 \\  
6707.596  & Cr I   & 4.208 & -2.667 \\  
6707.6453 & CN     & 0.946 & -3.330 \\  
6707.660  & C$_2$  & 0.926 & -1.743 \\  
6707.7561 & $^7$Li & 0.000 & -0.428 \\  
6707.7682 & $^7$Li & 0.000 & -0.206 \\  
6707.809  & CN     & 1.221 & -1.935 \\  
6707.8475 & CN     & 3.60  & -2.417 \\  
6707.8992 & CN     & 3.36  & -3.110 \\  
6707.9066 & $^7$Li & 0.000 & -1.509 \\  
6707.9080 & $^7$Li & 0.000 & -0.807 \\  
6707.9187 & $^7$Li & 0.000 & -0.807 \\  
6707.9196 & $^6$Li & 0.000 & -0.479 \\  
6707.9200 & $^7$Li & 0.000 & -0.807 \\  
6707.9230 & $^6$Li & 0.000 & -0.178 \\  
6707.9300 & CN     & 1.98  & -1.651 \\  
6707.970  & C$_2$  & 0.920 & -1.771 \\  
6707.980  & CN     & 2.372 & -3.527 \\  
6708.023  & Si I   & 6.00  & -2.80  \\  
6708.0261 & CN     & 1.98  & -2.031 \\  
6708.0728 & $^6$Li & 0.000 & -0.303 \\  
6708.094  & V I    & 1.218 & -2.922 \\  
6708.099  & Ce II  & 0.701 & -2.120 \\  
6708.1470 & CN     & 1.87  & -1.884 \\  
6708.282  & Fe I   & 4.988 & -2.70  \\  
6708.3146 & CN     & 2.64  & -1.719 \\  
6708.347  & Fe I   & 5.486 & -2.58  \\  
6708.3700 & CN     & 2.64  & -2.540 \\  
6708.420  & CN     & 0.768 & -3.358 \\  
6708.534  & Fe I   & 5.558 & -2.936 \\  
6708.5407 & CN     & 2.50  & -1.876 \\  
6708.577  & Fe I   & 5.446 & -2.684 \\  
\hline       
\end{tabular}
\end{table}

\clearpage

\scriptsize
\begin{table}
\caption{Stellar abundances [X/H] in LTE and NLTE, and errors due to uncertainties in the stellar parameters.}
\label{abund}
\centering 
\renewcommand{\footnoterule}{}  
\begin{tabular}{lrrrrrrrrrr} 
\hline    
\hline 
{Element}& LTE   & NLTE  & $\Delta \tef$ & $\Delta$log $g$ & $\Delta v_t$ & $\Delta$[Fe/H] & param\tablefootmark{a} & obs\tablefootmark{b} & total\tablefootmark{c} \\
{}       &       &       & +7K           &  +0.02 dex      & +0.01 km s$^{-1}$  & +0.01 dex   &  &  &  \\
{}       & (dex) & (dex) & (dex)         & (dex)           & (dex)       & (dex)        & (dex) & (dex) & (dex) \\
\hline
C  & 0.007 & 0.007 & -0.004 &  0.003 &  0.000 &  0.000 & 0.005 & 0.006 & 0.008\\  
O  & 0.012 & 0.011 & -0.005 &  0.001 & -0.001 &  0.002 & 0.006 & 0.008 & 0.010\\ 
Na & 0.014 & 0.014 &  0.004 &  0.000 &  0.000 & -0.001 & 0.004 & 0.006 & 0.007\\  
Mg & 0.013 & 0.012 &  0.004 & -0.002 & -0.001 & -0.001 & 0.005 & 0.005 & 0.007\\  
Al & 0.011 & 0.010 &  0.004 &  0.000 &  0.000 & -0.001 & 0.004 & 0.008 & 0.009\\  
Si & 0.022 &       &  0.002 &  0.001 & -0.001 &  0.001 & 0.003 & 0.003 & 0.004\\  
S  & 0.004 &       & -0.003 &  0.003 &  0.000 &  0.001 & 0.004 & 0.005 & 0.007\\  
K  & 0.007 & 0.007 &  0.006 & -0.007 & -0.002 &  0.001 & 0.009 & 0.009 & 0.013\\  
Ca & 0.024 & 0.023 &  0.005 & -0.002 & -0.002 &  0.000 & 0.006 & 0.003 & 0.006\\  
Sc & 0.025 &       &  0.000 &  0.007 & -0.001 &  0.003 & 0.008 & 0.004 & 0.009\\  
Ti & 0.018 & 0.017 &  0.007 &  0.001 & -0.002 & -0.001 & 0.007 & 0.002 & 0.008\\  
V  & 0.033 &       &  0.008 &  0.002 & -0.001 & -0.001 & 0.008 & 0.004 & 0.009\\  
Cr & 0.015 & 0.012 &  0.005 &  0.000 & -0.002 & -0.001 & 0.005 & 0.003 & 0.006\\  
Mn & 0.021 & 0.016 &  0.006 & -0.003 & -0.003 &  0.000 & 0.007 & 0.005 & 0.009\\  
Fe & 0.020 & 0.021 &  0.006 &  0.000 & -0.002 &  0.000 & 0.006 & 0.001 & 0.006\\  
Co & 0.026 & 0.024 &  0.005 &  0.002 & -0.001 &  0.000 & 0.005 & 0.003 & 0.006\\  
Ni & 0.025 &       &  0.004 &  0.000 & -0.002 &  0.000 & 0.004 & 0.001 & 0.005\\  
Cu & 0.014 &       &  0.005 &  0.000 & -0.002 &  0.000 & 0.005 & 0.005 & 0.007\\  
Zn & 0.019 & 0.018 &  0.001 &  0.001 & -0.003 &  0.002 & 0.004 & 0.005 & 0.006\\  
Y  & 0.021 &       &  0.001 &  0.007 & -0.004 &  0.003 & 0.009 & 0.004 & 0.010\\  
Zr & 0.041 & 0.041 &  0.002 &  0.008 & -0.002 &  0.003 & 0.009 & 0.005 & 0.010\\  
Ba & 0.024 & 0.023 &  0.002 &  0.002 & -0.004 &  0.005 & 0.007 & 0.005 & 0.009\\  
\hline       
\end{tabular}
\tablefoot{
\tablefoottext{a}{Adding errors in stellar parameters}
\tablefoottext{b}{Observational errors}
\tablefoottext{c}{Total error (stellar parameters and observational)}
}
\end{table}

\clearpage

\scriptsize
\onllongtab{3}{
\begin{longtable}{rrrrrrrrrrrrrrrrrrrrrrrrrrrr}
\caption{Adopted atomic data, equivalent widths, and differential NLTE corrections (HIP 56948 - Sun)} \\
\hline
\hline 
{Wavelength} & ion & $\chi_{exc}$ & log $gf$ & $C_6$ & EW & EW & $\Delta$NLTE \\
\hline
 (\AA) &     & (eV) &     &   & HIP 56948 & Sun & (dex) \\
\hline
\endfirsthead
\caption{Continued.} \\
\hline
{Wavelength} & ion & $\chi_{exc}$ & log $gf$ & $C_6$ & EW & EW & $\Delta$NLTE \\
\hline
 (\AA) &     & (eV) &     &   & HIP 56948 & Sun & (dex) \\
\hline
\endhead
\hline
\endfoot
\hline
\endlastfoot
 4445.471 & 26.00& 0.087  & -5.441 &   2.80    &  40.1 &  40.8 &  0.001\\
 5044.211 & 26.00& 2.8512 & -2.058 & 0.271E-30 &  73.7 &  73.6 &  0.001\\
 5225.525 & 26.00& 0.1101 & -4.789 & 0.123E-31 &  71.2 &  70.4 &  0.000\\
 5247.050 & 26.00& 0.0872 & -4.946 & 0.122E-31 &  66.0 &  66.3 &  0.001\\
 5250.208 & 26.00& 0.1212 & -4.938 & 0.123E-31 &  64.7 &  64.9 &  0.001\\
 5651.469 & 26.00& 4.473  & -1.75  & 0.483E-30 &  19.3 &  18.9 &  0.001\\
 5661.348 & 26.00& 4.2843 & -1.756 & 0.324E-30 &  23.7 &  22.8 &  0.001\\
 5679.023 & 26.00& 4.652  & -0.75  & 0.813E-30 &  59.9 &  59.4 &  0.000\\
 5696.089 & 26.00& 4.548  & -1.720 & 0.578E-30 &  14.4 &  13.9 &  0.000\\
 5701.544 & 26.00& 2.559  & -2.216 & 0.495E-31 &  85.7 &  84.2 &  0.001\\
 5705.464 & 26.00& 4.301  & -1.355 & 0.302E-30 &  38.2 &  38.0 &  0.001\\
 5778.453 & 26.00& 2.588  & -3.430 & 0.495E-31 &  22.6 &  22.5 &  0.001\\
 5784.658 & 26.00& 3.396  & -2.532 & 0.357E-30 &  27.1 &  27.1 &  0.000\\
 5793.914 & 26.00& 4.220  & -1.619 & 0.272E-30 &  34.6 &  33.9 &  0.000\\
 5809.218 & 26.00& 3.883  & -1.609 & 0.565E-30 &  51.9 &  51.2 &  0.000\\
 5855.076 & 26.00& 4.6075 & -1.478 & 0.574E-30 &  23.8 &  23.1 &  0.000\\
 5916.247 & 26.00& 2.453  & -2.936 & 0.429E-31 &  56.5 &  55.8 &  0.001\\
 5956.694 & 26.00& 0.8589 & -4.605 & 0.155E-31 &  50.8 &  50.9 &  0.001\\
 6027.050 & 26.00& 4.0758 & -1.09  &   2.80    &  64.7 &  63.9 &  0.001\\
 6065.482 & 26.00& 2.6085 & -1.530 & 0.471E-31 & 117.6 & 117.2 &  0.001\\
 6093.644 & 26.00& 4.607  & -1.30  & 0.441E-30 &  32.0 &  31.6 &  0.001\\ 
 6096.665 & 26.00& 3.9841 & -1.81  & 0.575E-30 &  38.1 &  37.1 &  0.000\\
 6151.618 & 26.00& 2.1759 & -3.299 & 0.255E-31 &  50.6 &  49.7 &  0.000\\
 6165.360 & 26.00& 4.1426 & -1.46  &   2.80    &  46.1 &  44.9 &  0.000\\
 6173.335 & 26.00& 2.223  & -2.880 & 0.265E-31 &  69.2 &  68.8 &  0.001\\
 6200.313 & 26.00& 2.6085 & -2.437 & 0.458E-31 &  75.1 &  74.3 &  0.000\\
 6213.430 & 26.00& 2.2227 & -2.52  & 0.262E-31 &  83.5 &  82.3 &  0.000\\
 6219.281 & 26.00& 2.198  & -2.433 & 0.258E-31 &  90.3 &  89.0 &  0.001\\
 6240.646 & 26.00& 2.2227 & -3.233 & 0.314E-31 &  49.2 &  49.4 &  0.001\\
 6252.555 & 26.00& 2.4040 & -1.687 & 0.384E-31 & 121.3 & 120.7 &  0.001\\
 6265.134 & 26.00& 2.1759 & -2.550 & 0.248E-31 &  84.9 &  85.2 &  0.001\\
 6270.225 & 26.00& 2.8580 & -2.54  & 0.458E-31 &  51.4 &  51.4 &  0.001\\
 6430.846 & 26.00& 2.1759 & -2.006 & 0.242E-31 & 112.6 & 111.2 &  0.001\\
 6498.939 & 26.00& 0.9581 & -4.699 & 0.153E-31 &  46.9 &  46.4 &  0.000\\
 6593.871 & 26.00& 2.4326 & -2.422 & 0.369E-31 &  83.5 &  83.6 &  0.001\\
 6703.567 & 26.00& 2.7585 & -3.023 & 0.366E-31 &  37.6 &  37.5 &  0.001\\
 6705.102 & 26.00& 4.607  & -0.98  &   2.80    &  47.3 &  47.6 &  0.001\\
 6713.745 & 26.00& 4.795  & -1.40  & 0.430E-30 &  21.8 &  21.4 &  0.000\\
 6726.667 & 26.00& 4.607  & -1.03  & 0.482E-30 &  47.7 &  47.5 &  0.001\\
 6750.152 & 26.00& 2.4241 & -2.621 & 0.411E-31 &  74.2 &  74.0 &  0.001\\
 6810.263 & 26.00& 4.607  & -0.986 & 0.450E-30 &  51.0 &  50.3 &  0.000\\
 6837.006 & 26.00& 4.593  & -1.687 & 0.246E-31 &  18.1 &  18.2 &  0.001\\
 4508.288 & 26.10& 2.8557 & -2.44  & 0.956E-32 &  84.8 &  84.2 & \\
 4520.224 & 26.10& 2.8068 & -2.65  & 0.857E-32 &  81.1 &  81.1 & \\
 4576.340 & 26.10& 2.8443 & -2.95  & 0.943E-32 &  62.5 &  62.3 & \\
 4620.521 & 26.10& 2.8283 & -3.21  & 0.930E-32 &  53.6 &  53.9 & \\
 5197.577 & 26.10& 3.2306 & -2.22  & 0.869E-32 &  81.0 &  80.0 & \\
 5234.625 & 26.10& 3.2215 & -2.18  & 0.869E-32 &  83.1 &  82.5 & \\
 5264.812 & 26.10& 3.2304 & -3.13  & 0.943E-32 &  44.4 &  44.4 & \\
 5414.073 & 26.10& 3.2215 & -3.58  & 0.930E-32 &  28.1 &  27.5 & \\
 5425.257 & 26.10& 3.1996 & -3.22  & 0.845E-32 &  40.9 &  41.7 & \\
 6369.462 & 26.10& 2.8912 & -4.11  & 0.742E-32 &  19.7 &  19.1 & \\
 6432.680 & 26.10& 2.8912 & -3.57  & 0.742E-32 &  42.8 &  41.2 & \\
 7711.724 & 26.10& 3.9034 & -2.50  & 0.930E-32 &  47.4 &  47.3 & \\
 5052.167 & 06.0 & 7.685  & -1.24  &   2.80    &  34.4 &  33.0 & -0.001 \\
 5380.337 & 06.0 & 7.685  & -1.57  &   2.80    &  22.3 &  21.7 &  0.000 \\
 6587.61  & 06.0 & 8.537  & -1.05  &   2.80    &  14.1 &  13.9 &  0.000 \\
 7111.47  & 06.0 & 8.640  & -1.07  & 0.291E-29 &  11.0 &  11.2 &  0.000 \\
 7113.179 & 06.0 & 8.647  & -0.76  & 0.297E-29 &  22.5 &  22.1 &  0.000 \\
 7771.944 & 08.0 & 9.146  &  0.37  & 0.841E-31 &  70.7 &  68.7 &  0.000 \\
 7774.166 & 08.0 & 9.146  &  0.22  & 0.841E-31 &  61.2 &  59.8 & -0.001 \\
 7775.388 & 08.0 & 9.146  &  0.00  & 0.841E-31 &  49.1 &  48.8 & -0.002 \\
 4751.822 & 11.0 & 2.1044 & -2.078 &   2.80    &  13.5 &  13.7 &  0.000\\
 5148.838 & 11.0 & 2.1023 & -2.044 &   2.80    &  10.9 &  10.9 &  0.001\\
 6154.225 & 11.0 & 2.1023 & -1.547 &   2.80    &  39.1 &  38.3 & -0.002\\
 6160.747 & 11.0 & 2.1044 & -1.246 &   2.80    &  55.5 &  54.0 &  0.002\\
 5711.088 & 12.0 & 4.345  & -1.729 &   2.80    & 106.1 & 105.3 & -0.002\\
 6318.717 & 12.0 & 5.108  & -1.945 &   2.80    &  38.6 &  38.3 &  0.000\\
 6319.236 & 12.0 & 5.108  & -2.165 &   2.80    &  28.5 &  28.3 &  0.001\\
 6696.018 & 13.0 & 3.143  & -1.481 &   2.80    &  40.6 &  40.4 &  0.000\\
 6698.667 & 13.0 & 3.143  & -1.782 &   2.80    &  22.5 &  22.3 & -0.002\\
 5488.983 & 14.0 & 5.614  & -1.69  &   2.80    &  20.6 &  20.4 & \\
 5645.611 & 14.0 & 4.929  & -2.04  &   2.80    &  37.0 &  35.6 & \\
 5684.484 & 14.0 & 4.953  & -1.55  &   2.80    &  63.6 &  62.1 & \\
 5690.425 & 14.0 & 4.929  & -1.77  &   2.80    &  49.8 &  48.9 & \\
 5701.104 & 14.0 & 4.930  & -1.95  &   2.80    &  40.1 &  38.9 & \\
 5793.073 & 14.0 & 4.929  & -1.96  &   2.80    &  44.7 &  44.4 & \\
 6125.021 & 14.0 & 5.614  & -1.50  &   2.80    &  33.3 &  32.0 & \\
 6145.015 & 14.0 & 5.616  & -1.41  &   2.80    &  41.0 &  39.1 & \\
 6243.823 & 14.0 & 5.616  & -1.27  &   2.80    &  47.5 &  46.8 & \\
 6244.476 & 14.0 & 5.616  & -1.32  &   2.80    &  48.6 &  47.4 & \\
 6721.848 & 14.0 & 5.862  & -1.12  &   2.80    &  46.1 &  45.4 & \\
 6741.63  & 14.0 & 5.984  & -1.65  &   2.80    &  16.9 &  16.8 & \\
 6046.000 & 16.0 & 7.868  & -0.15  &   2.80    &  19.4 &  19.5 & \\
 6052.656 & 16.0 & 7.870  & -0.4   &   2.80    &  12.7 &  12.7 & \\
 6743.54  & 16.0 & 7.866  & -0.6   &   2.80    &   9.4 &   9.1 & \\
 6757.153 & 16.0 & 7.870  & -0.15  &   2.80    &  18.8 &  18.3 & \\
 7698.974 & 19.0 & 0.000  & -0.168 & 0.104E-30 & 158.3 & 158.6 &  0.000\\
 4512.268 & 20.0 & 2.526  & -1.901 &   2.80    &  23.3 &  22.5 &       \\
 5260.387 & 20.0 & 2.521  & -1.719 & 0.727E-31 &  32.6 &  32.6 &       \\
 5512.980 & 20.0 & 2.933  & -0.464 &   2.80    &  89.1 &  86.9 &  0.001\\
 5590.114 & 20.0 & 2.521  & -0.571 & 0.636E-31 &  92.5 &  91.3 & -0.001\\
 5867.562 & 20.0 & 2.933  & -1.57  &   2.80    &  24.3 &  24.2 &  0.000\\
 6166.439 & 20.0 & 2.521  & -1.142 & 0.595E-30 &  70.4 &  70.1 & -0.001\\
 6169.042 & 20.0 & 2.523  & -0.797 & 0.595E-30 &  93.3 &  92.4 &  0.000\\
 6455.598 & 20.0 & 2.523  & -1.34  & 0.509E-31 &  57.4 &  56.5 &       \\
 6471.662 & 20.0 & 2.525  & -0.686 & 0.509E-31 &  92.8 &  92.2 &  0.000\\
 6499.650 & 20.0 & 2.523  & -0.818 & 0.505E-31 &  87.2 &  85.4 & -0.001\\
 6798.470 & 20.0 & 2.709  & -2.45  &   2.80    &   7.7 &   7.3 & \\
 4743.821 & 21.0 & 1.4478 &  0.35  & 0.597E-31 &   9.7 &   9.8 & \\
 5081.57  & 21.0 & 1.4478 &  0.30  &   2.80    &   7.9 &   8.2 & \\
 5520.497 & 21.0 & 1.8649 &  0.55  &   2.80    &   6.8 &   6.8 & \\
 5671.821 & 21.0 & 1.4478 &  0.55  &   2.80    &  15.6 &  15.3 & \\
 4420.661 & 21.1 & 0.6184 & -2.273 &   2.80    &  16.2 &  16.5 & \\
 5657.87  & 21.1 & 1.507  & -0.30  &   2.80    &  69.4 &  67.3 & \\
 5684.19  & 21.1 & 1.507  & -0.95  &   2.80    &  39.5 &  38.2 & \\
 6245.63  & 21.1 & 1.507  & -1.030 &   2.80    &  34.2 &  33.8 & \\
 6279.76  & 21.1 & 1.500  & -1.2   &   2.80    &  30.8 &  29.8 & \\
 6300.698 & 21.1 & 1.507  & -2.0   &   2.80    &   5.9 &   5.6 & \\
 6320.843 & 21.1 & 1.500  & -1.85  &   2.80    &   9.9 &   9.3 & \\
 6604.578 & 21.1 & 1.3569 & -1.15  &   2.80    &  37.1 &  36.6 & \\
 4281.369 & 22.0 & 0.8129 & -1.359 & 0.502E-31 &  24.6 &  24.4 & -0.001\\
 4465.802 & 22.0 & 1.7393 & -0.163 & 0.398E-31 &  38.1 &  37.9 & -0.001\\
 4758.120 & 22.0 & 2.2492 &  0.425 & 0.384E-31 &  42.6 &  42.9 & -0.001\\
 4759.272 & 22.0 & 2.2555 &  0.514 & 0.386E-31 &  46.4 &  46.3 &  0.000\\
 5022.871 & 22.0 & 0.8258 & -0.434 & 0.358E-31 &  70.1 &  69.8 &  0.002\\
 5113.448 & 22.0 & 1.443  & -0.783 & 0.306E-31 &  27.3 &  27.6 & -0.002\\
 5219.700 & 22.0 & 0.021  & -2.292 & 0.208E-31 &  29.3 &  29.0 & -0.002\\
 5490.150 & 22.0 & 1.460  & -0.933 & 0.541E-31 &  21.6 &  21.4 & -0.003\\
 5866.452 & 22.0 & 1.066  & -0.840 & 0.216E-31 &  47.5 &  47.9 &  0.002\\
 6126.217 & 22.0 & 1.066  & -1.424 & 0.206E-31 &  22.1 &  22.2 & -0.001\\
 6258.104 & 22.0 & 1.443  & -0.355 & 0.481E-31 &  51.9 &  51.4 & -0.003\\
 6261.101 & 22.0 & 1.429  & -0.479 & 0.468E-31 &  49.5 &  48.7 & -0.003\\
 4583.408 & 22.1 & 1.165  & -2.87  &   2.80    &  31.3 &  31.4 &  0.000\\
 4636.33  & 22.1 & 1.16   & -3.152 &   2.80    &  19.2 &  19.1 &  0.000\\
 4657.212 & 22.1 & 1.243  & -2.8   &   2.80    &  32.4 &  32.2 &  0.000\\
 4865.611 & 22.1 & 1.116  & -2.81  &   2.80    &  40.3 &  39.3 &  0.000\\
 4911.193 & 22.1 & 3.123  & -0.537 &   2.80    &  51.6 &  51.8 & -0.002\\
 5211.54  & 22.1 & 2.59   & -1.49  &   2.80    &  34.2 &  33.0 & -0.001\\
 5381.015 & 22.1 & 1.565  & -1.97  &   2.80    &  61.1 &  59.9 &  0.000\\
 5418.767 & 22.1 & 1.582  & -2.11  &   2.80    &  49.5 &  48.8 &  0.001\\
 4594.119 & 23.0 & 0.068  & -0.67  & 0.216E-31 &  55.6 &  55.8 & \\
 4875.486 & 23.0 & 0.040  & -0.81  & 0.198E-31 &  45.9 &  45.4 & \\
 5670.85  & 23.0 & 1.080  & -0.42  & 0.358E-31 &  19.5 &  19.2 & \\
 5727.046 & 23.0 & 1.081  & -0.011 & 0.435E-31 &  39.4 &  38.9 & \\
 6039.73  & 23.0 & 1.063  & -0.65  & 0.398E-31 &  13.8 &  13.1 & \\
 6081.44  & 23.0 & 1.051  & -0.578 & 0.389E-31 &  14.7 &  14.4 & \\
 6090.21  & 23.0 & 1.080  & -0.062 & 0.398E-31 &  33.8 &  33.1 & \\
 6119.528 & 23.0 & 1.064  & -0.320 & 0.389E-31 &  23.4 &  22.6 & \\
 6199.20  & 23.0 & 0.286  & -1.28  & 0.196E-31 &  14.6 &  14.6 & \\
 6251.82  & 23.0 & 0.286  & -1.34  & 0.196E-31 &  16.2 &  15.3 & \\
 6274.65  & 23.0 & 0.267  & -1.67  & 0.194E-31 &   9.2 &   8.5 & \\
 6285.160 & 23.0 & 0.275  & -1.51  & 0.194E-31 &  11.2 &  10.8 & \\
 4801.047 & 24.0 & 3.1216 & -0.130 & 0.452E-31 &  49.2 &  49.4 & -0.003\\
 4936.335 & 24.0 & 3.1128 & -0.25  & 0.432E-31 &  44.3 &  44.3 & -0.003\\
 5238.964 & 24.0 & 2.709  & -1.27  & 0.519E-31 &  16.5 &  16.3 & -0.002\\
 5247.566 & 24.0 & 0.960  & -1.59  & 0.392E-31 &  81.3 &  81.1 & -0.003\\
 5272.007 & 24.0 & 3.449  & -0.42  & 0.315E-30 &  24.6 &  24.0 & -0.003\\
 5287.20  & 24.0 & 3.438  & -0.87  & 0.309E-30 &  11.4 &  11.5 & -0.003\\
 5783.09  & 24.0 & 3.323  & -0.43  & 0.802E-30 &  31.8 &  31.2 & -0.002\\
 4588.199 & 24.1 & 4.071  & -0.594 &   2.80    &  70.8 &  69.7 & -0.003\\
 4592.049 & 24.1 & 4.073  & -1.252 &   2.80    &  47.5 &  47.7 & -0.002\\
 5237.328 & 24.1 & 4.073  & -1.087 &   2.80    &  54.0 &  54.0 &  0.000\\
 5246.767 & 24.1 & 3.714  & -2.436 &   2.80    &  16.1 &  15.5 &  0.001\\
 5502.067 & 24.1 & 4.168  & -2.049 &   2.80    &  19.3 &  18.7 &  0.000\\
 4082.939 & 25.0 & 2.1782 & -0.354 & 0.255E-31 &  90.0 &  89.8 & -0.007\\
 4739.10  & 25.0 & 2.9408 & -0.490 & 0.352E-31 &  61.5 &  60.3 & -0.002\\
 5004.891 & 25.0 & 2.9197 & -1.63  & 0.314E-31 &  14.0 &  14.1 & -0.004\\
 6013.49  & 25.0 & 3.073  & -0.251 &   2.80    &  87.8 &  86.5 & -0.005\\
 6016.64  & 25.0 & 3.073  & -0.084 &   2.80    &  97.0 &  96.1 & -0.006\\
 6021.79  & 25.0 & 3.076  & +0.034 &   2.80    &  89.5 &  89.3 & -0.005\\
 5212.691 & 27.0 & 3.5144 & -0.11  & 0.339E-30 &  21.4 &  20.6 & -0.002\\
 5247.911 & 27.0 & 1.785  & -2.08  & 0.327E-31 &  18.1 &  17.7 & -0.001\\
 5483.352 & 27.0 & 1.7104 & -1.49  & 0.289E-31 &  51.1 &  51.2 & -0.002\\
 5530.774 & 27.0 & 1.710  & -2.23  & 0.226E-31 &  18.9 &  19.1 & -0.002\\
 5647.23  & 27.0 & 2.280  & -1.56  & 0.414E-31 &  15.0 &  14.7 & -0.002\\
 6189.00  & 27.0 & 1.710  & -2.46  & 0.206E-31 &  11.6 &  11.3 & -0.002\\
 6454.995 & 27.0 & 3.6320 & -0.25  & 0.378E-30 &  14.7 &  14.1 & -0.002\\
 5589.358 & 28.0 & 3.898  & -1.14  & 0.398E-30 &  27.6 &  27.0 & \\
 5643.078 & 28.0 & 4.164  & -1.25  & 0.379E-30 &  16.1 &  15.7 & \\
 6086.282 & 28.0 & 4.266  & -0.51  & 0.406E-30 &  45.7 &  44.9 & \\
 6108.116 & 28.0 & 1.676  & -2.44  & 0.248E-31 &  64.3 &  63.3 & \\
 6130.135 & 28.0 & 4.266  & -0.96  & 0.391E-30 &  24.4 &  23.6 & \\
 6204.604 & 28.0 & 4.088  & -1.14  & 0.277E-30 &  23.1 &  22.4 & \\
 6223.984 & 28.0 & 4.105  & -0.98  & 0.393E-30 &  28.7 &  28.3 & \\
 6378.25  & 28.0 & 4.1535 & -0.90  & 0.391E-30 &  32.5 &  31.8 & \\
 6767.772 & 28.0 & 1.826  & -2.17  &   2.80    &  79.9 &  79.2 & \\
 6772.315 & 28.0 & 3.657  & -0.99  & 0.356E-30 &  50.6 &  50.0 & \\
 7727.624 & 28.0 & 3.678  & -0.4   & 0.343E-30 &  91.4 &  90.7 & \\
 7788.930 & 28.0 & 1.950  & -2.0   & 0.218E-31 &  92.7 &  92.1 & \\
 7797.586 & 28.0 & 3.89   & -0.34  &   2.80    &  80.0 &  78.6 & \\
 5105.541 & 29.0 & 1.39   & -1.516 &   2.80    &  91.5 &  91.5 & \\
 5218.197 & 29.0 & 3.816  &  0.476 &   2.80    &  52.2 &  51.2 & \\
 5220.066 & 29.0 & 3.816  & -0.448 &   2.80    &  17.3 &  17.3 & \\
 7933.13  & 29.0 & 3.79   & -0.368 &   2.80    &  30.5 &  30.8 & \\
 4722.159 & 30.0 & 4.03   & -0.38  &   2.80    &  71.9 &  71.2 & -0.001\\
 4810.534 & 30.0 & 4.08   & -0.16  &   2.80    &  73.8 &  73.3 & -0.001\\
 6362.35  & 30.0 & 5.79   &  0.14  &   2.80    &  21.6 &  21.1 &  0.000\\
 4854.867 & 39.1 & 0.9923 & -0.38  &   2.80    &  48.3 &  48.4 & \\
 4883.685 & 39.1 & 1.0841 &  0.07  &   2.80    &  57.6 &  57.1 & \\
 4900.110 & 39.1 & 1.0326 & -0.09  &   2.80    &  55.5 &  55.3 & \\
 5087.420 & 39.1 & 1.0841 & -0.17  &   2.80    &  49.1 &  48.4 & \\
 5200.413 & 39.1 & 0.9923 & -0.57  &   2.80    &  39.0 &  38.6 & \\
 4050.320 & 40.1 & 0.713  & -1.06  &   2.80    &  23.8 &  23.4 &  0.001\\
 4208.980 & 40.1 & 0.713  & -0.51  &   2.80    &  43.2 &  42.0 &  0.000\\
 4442.992 & 40.1 & 1.486  & -0.42  &   2.80    &  25.7 &  24.7 &  0.000\\
 5853.67  & 56.1 & 0.604  & -0.91  & 0.53E-31  &  64.6 &  63.7 & -0.002\\
 6141.71  & 56.1 & 0.704  & -0.08  & 0.53E-31  & 116.2 & 115.7 &  0.000\\
 6496.90  & 56.1 & 0.604  & -0.38  & 0.53E-31  &  99.5 &  99.2 &  0.000\\
\label{tabew}
\end{longtable}
}

\end{document}